\documentclass[preprint,12pt,compress,hyphens]{elsarticle}

\usepackage{subcaption}
\usepackage{lineno}
\usepackage{algorithm}
\usepackage{algpseudocode}
\usepackage{amsmath}
\usepackage{amssymb}
\usepackage{array}
\usepackage{booktabs}
\usepackage{multirow}
\usepackage{makecell}
\usepackage{tabularx}
\usepackage{geometry}
\usepackage{caption}
\usepackage[table, svgnames, dvipsnames]{xcolor}
\usepackage[figuresright]{rotating}
\usepackage[hidelinks]{hyperref}
\hypersetup{breaklinks=true}

\DeclareMathOperator\erf{erf}

\newtheorem{definition}{Definition}
\newtheorem{remark}{Remark}
\newtheorem{assumption}{Assumption}

\makeatletter
\def\ps@pprintTitle{%
\let\@oddhead\@empty
\let\@evenhead\@empty
\let\@oddfoot\@empty
\let\@evenfoot\@oddfoot
}
\makeatother

\def\tblwidth{\linewidth}
\newcolumntype{L}{@{\extracolsep{\fill}}l}

\begin{document}

\begin{frontmatter}

\title{Vector-level Feedforward Control of LPBF Melt Pool Area Using a Physics-Based Thermal Model}  
\author[UM]{Nicholas Kirschbaum}\ead{nkirsch@umich.edu}
\author[UM]{Nathaniel Wood}\ead{nathaniel.wood.2.ctr@afrl.af.mil}
\author[LLNL]{Chang-Eun Kim}\ead{kim87@llnl.gov}
\author[LLNL]{Thejaswi U. Tumkur}\ead{tumkurumanat1@llnl.gov}
\author[UM]{Chinedum Okwudire\corref{cor1}}\ead{okwudire@umich.edu}
\affiliation[UM]{organization={Department of Mechanical Engineering, University of Michigan},
            city={Ann Arbor},
            postcode={48109}, 
            state={MI},
            country={USA}}
\affiliation[LLNL]{organization={Lawrence Livermore National Laboratory},
            city={Livermore},
            postcode={94550}, 
            state={CA},
            country={USA}}
\cortext[cor1]{Corresponding author}

\begin{abstract}
Laser powder bed fusion (LPBF) is an additive manufacturing technique that has gained popularity thanks to its ability to produce geometrically complex, fully dense metal parts. However, these parts are prone to internal defects and geometric inaccuracies, stemming in part from variations in the melt pool. This paper proposes a novel vector-level feedforward control framework for regulating melt pool area in LPBF. By decoupling part-scale thermal behavior from small-scale melt pool physics, the controller provides a scale-agnostic prediction of melt pool area and efficient optimization over it. This is done by operating on two coupled lightweight models: a finite-difference thermal model that efficiently captures vector-level temperature fields and a reduced-order, analytical melt pool model. Each model is calibrated separately with minimal single-track and 2D experiments, and the framework is validated on a complex 3D geometry in both Inconel 718 and 316L stainless steel. Results showed that feedforward vector-level laser power scheduling reduced geometric inaccuracy in key dimensions by \(62\%\), overall porosity by \(16.5\%\), and photodiode variation by \(6.8\%\) on average. Overall, this modular, data-efficient approach demonstrates that proactively compensating for known thermal effects can significantly improve part quality while remaining computationally efficient and readily extensible to other materials and machines.
\end{abstract}

\end{frontmatter}

\pagebreak
\section{Introduction}\label{text_intro}

Laser powder bed fusion (LPBF) has gained significant traction as an additive manufacturing (AM) method in high-tech industries, including aerospace, automotive, and biomedical engineering. By selectively fusing layers of metal powder, LPBF enables the fabrication of complex geometries, reduces material waste, and supports cost-effective, low-volume production \cite{ngoAdditiveManufacturing3D2018,narasimharajuComprehensiveReviewLaser2022,chowdhuryLaserPowderBed2022}. Despite these advantages, the process is susceptible to defects that can compromise the mechanical properties of finished parts and, in severe cases, lead to print failures \cite{chowdhuryLaserPowderBed2022, zhaiAdditiveManufacturingMaking2014}. Minimizing these defects is, therefore, a key challenge in LPBF.

Central to these defects in LPBF is the variability of the melt pool. These variations in the amount of molten material can lead to defects such as balling, dross, and pore formation via keyholing or lack of fusion \cite{grassoProcessDefectsSitu2017,wangUnderstandingMeltPool2023}. These defects adversely affect mechanical properties, underscoring the importance of melt pool regulation \cite{rubenchikScalingLawsAdditive2018}. However, the melt pool's small size, orders of magnitude smaller than the final part, and the speed at which processing happens, up to meters per second, make real-time monitoring and control exceedingly difficult \cite{mazzoleniRealTimeObservationMelt2020,maniReviewMeasurementScience2017}. 
Additional uncertainty arises from the numerous factors that influence melt pool behavior, including powder properties, laser parameters, environmental conditions, and underlying geometry. These multi-scale effects make melt pool regulation an ongoing challenge \cite{narasimharajuComprehensiveReviewLaser2022,chowdhuryLaserPowderBed2022,laneThermographicMeasurementsCommercial2016}. Currently, process parameters are chosen through the design of process maps from a combination of single-track scans and simple test geometries \cite{seifiOverviewMaterialsQualification2016}. However, for more complex geometries, these process maps are insufficient, so parameters are optimized by trial and error, which is time-consuming and expensive \cite{knappBuildingBlocksDigital2017}.

In response to these challenges, a common strategy is to adjust the laser power per layer, much like an experienced technician would do manually. For example, \cite{rezaeifarOnlineMeltPool2021,kavasLayertolayerClosedloopFeedback2023,rienscheRapidAutonomousShapeAgnostic2025} developed controllers that automatically modulated the power for each layer based on measurements of the previous layer \cite{rezaeifarOnlineMeltPool2021,kavasLayertolayerClosedloopFeedback2023} or model-based predictions of the layer's temperatures \cite{rienscheRapidAutonomousShapeAgnostic2025}. Additionally, commercial systems, such as Smart Fusion from EOS \cite{yagmurHitchhikersGuideSmartFusion2023}, use optical tomography to adjust the energy input at every layer. Measurements are only acted on once per layer, and models typically feature simplified physics that only update once per layer. While effective in reducing layer-to-layer thermal variations, these methods cannot mitigate intra-layer fluctuations that arise in geometrically complex builds.

Motivated by the need for tighter process control, researchers have explored real-time laser power feedback strategies. Compared to layerwise control, laser power is adjusted much more frequently, often multiple times within a vector, that is, a single line that the laser marks. For instance, \cite{craeghsFeedbackControlLayerwise2010,renkenInprocessClosedloopControl2019,hussainFeedbackControlMelt2021,shkorutaRealTimeImageBasedFeedback2021,wangRealtimeProcessMonitoring2023} all developed various PID controllers based on photodiode signal \cite{craeghsFeedbackControlLayerwise2010,wangRealtimeProcessMonitoring2023}, coaxial camera signal \cite{shkorutaRealTimeImageBasedFeedback2021}, and simulated melt pool area \cite{hussainFeedbackControlMelt2021}. Additionally, photodiode-based feedback approaches have been implemented in some commercial systems, such as AconityCONTROL \cite{Aconity3DGmbHAdditive2025}. While all of these methods showed improvements, they remain fundamentally reactive and cannot account for known geometric variations (e.g., overhangs or shorter vectors) in advance. Moreover, changing process parameters in real-time, whether layerwise or more frequently, complicates certifying parts for critical applications \cite{evertonReviewInsituProcess2016}.

Driven by the limitations of purely reactive feedback strategies and the need for certification, researchers have explored feedforward control methods to anticipate and compensate for changes in advance. In one demonstration, Renken et al. \cite{renkenInprocessClosedloopControl2019} scheduled laser power for a few single tracks over an overhang based on temperatures from a high-fidelity simulation. The scan vector was subdivided into two regions: one for solid material and one for the overhang. The power decreased in the overhang, demonstrating how feedforward control can proactively compensate for known geometric variations. However, simulating the melt pool directly is too computationally complex for more than a few single tracks.

As such, much feedforward control research has investigated analytical or geometric methods. For example, Wang et al. \cite{wangModelbasedFeedforwardControl2020} derived an analytical model for the melt pool, accounting for prior scan vectors using superposition. The laser power was adjusted continuously in each vector to minimize overheating, and this adjustment successfully reduced variation. However, it was only tested on single layers, not for full 3D geometries. In another study, Yeung et al. \cite{yeungPartGeometryConductionbased2019} developed a geometric conductance factor, dependent on nearby solid material, to guide continuous power adjustments within each vector. These were promising, but limited. Analytical methods make significant approximations, making extending them to multiple layers difficult, and geometric methods provide simple approximations of complex thermal effects. 

To overcome these limitations, researchers have explored machine learning as a means to alleviate the computational burden of part-scale simulation and mitigate generalizability issues associated with analytical and geometric methods. These approaches utilize data-driven models to estimate melt pool behavior, aiming to enable robust full-scale control solutions. Lapointe et al. \cite{lapointePhotodiodebasedMachineLearning2022} trained a forward–inverse model that adjusted the laser power for each vector, demonstrating improved print accuracy but limited generalizability to new geometric features and required over 500,000 tracks to be scanned in 91 training prints. Similarly, Carter et al. \cite{carterMachineLearningGuided2024} used photodiode measurements and a temperature-to-power calibration to modify laser power, showing decreased porosity. However, their model was only tested on the same part for which it was trained. Conversely, Ren et al. \cite{renPhysicsInformedTwoLevelMachineLearning2021} developed a physics-informed two-level machine learning model that aimed to predict melt pool size by first predicting the prescan temperature using one model and training another model to predict the melt pool volume. This avoided the need to learn the entire relationship, but it required the prescan temperature to be measured for training. As such, this showed significant promise but was trained on simulated data and only used for prediction. Taken together, machine learning methods show promise in part-scale prediction and control; however, their significant data burden and issues with generalizability diminish their practicality. 

Thus, a key gap remains for an easily implementable part-scale method that does not require extensive calibration data, unlike current machine learning methods, but is generalizable and relatively computationally efficient, unlike current physics-based methods. To address these limitations, this paper presents a novel vector-level feedforward control framework that combines an analytical melt pool model with a part-scale physics-based thermal model. This approach decouples part-scale thermal effects from small-scale melt pool dynamics, offering a computationally efficient solution with minimal calibration requirements. Through this paper, we make the following original contributions to the literature:
\begin{enumerate}
    \item[(1)] We introduce a novel framework for vector-level feedforward control in LPBF that explicitly decouples part-scale thermal history from localized melt pool dynamics, enabling scalable, high-fidelity feedforward control across complex 3D geometries.
    \item[(2)] We develop a coupled analytical-numerical model that integrates a finite difference thermal simulation with an explicit analytical melt pool model, enabling closed-form vector-wise laser power scheduling. 
    \item[(3)] We propose a new, data-efficient calibration strategy for the melt pool model that, in addition to laser power and scan speed, uses controlled variations in baseplate temperature during single-track experiments to parameterize the thermal influence of subsurface conditions.
    
\end{enumerate}

We validate the feedforward control framework and coupled analytical-numerical model at the part scale by printing a complex test geometry using the feedforward control in two separate materials: Inconel 718 (IN718) and 316L stainless steel (316LSS). The results demonstrate reduced variation in the photodiode signal, improved geometric accuracy, and a decrease in defects compared to the nominal parameters.

The remainder of this paper is organized as follows. Section~\ref{text_modelAndControl} describes the modeling approach and controller framework, outlining how the part-scale and melt pool models are integrated. Section~\ref{text_modelCalibration} details the calibration procedure for both models using experimental data. Section~\ref{sec:validation_test_setup} explains the print setup and how we use the calibrated models to generate power schedules. We present and discuss the validation experiments in Section~\ref{text_results}, including multi-material testing. Finally, in Section~\ref{text_conclusion}, we summarize the key findings and suggest avenues for future research.

\section{Modeling and Controller Framework}\label{text_modelAndControl}

\begin{figure*}
    \centering
    \includegraphics[width=0.95\linewidth]{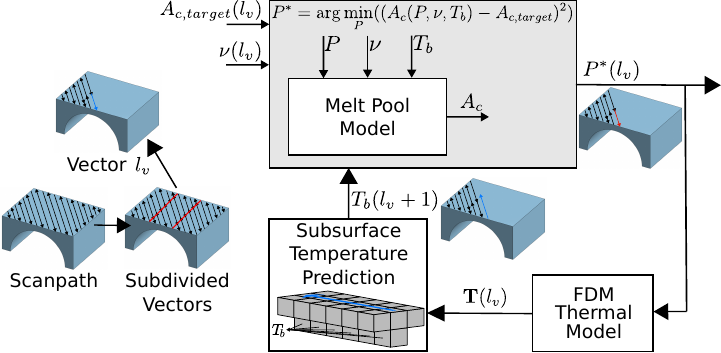}
    \caption{The overarching controller framework. First, the scanpath (left) is broken down into vectors and subdivided based on expected thermal conditions. Then, the scan speed \(\nu\), and melt pool area target \(A_{c,target}\) for vector \(l_v\) are fed into the optimizer along with subsurface temperature \(T_b\). The optimizer then finds an optimal laser power \(P^*\) to minimize the difference between \(A_{c,target}\) and predicted melt pool \(A_c\). Once \(P^*\) for vector \(l_v\) is found, it is fed into an FDM thermal model to simulate vector \(l_v\). Then, using the temperature field after the vector is simulated \(\mathbf{T}(l_{v})\), \(T_b\) for the next vector \(l_v+1\) is determined.}
    \label{fig:overaching_controller_design}
\end{figure*}

The proposed controller framework operates by coupling a fast, part-scale thermal model with a reduced-order melt pool model, as shown in Figure~\ref{fig:overaching_controller_design}. Both models are calibrated independently and then connected. The process is as follows: the scanpath is broken down into vectors. Then, for each vector, a finite difference method (FDM) thermal model predicts subsurface temperatures, capturing key geometric effects such as overhangs, thin walls, and scan path. Then, these subsurface temperature estimates, located underneath each vector, are passed to an analytical melt pool model, which computes the corresponding melt pool dimensions. By optimizing power inputs to this melt pool model, which is only feasible due to its lightweight nature, the controller generates a feedforward sequence of laser commands to regulate the melt pool area at the vector level. 

\subsection{Preliminaries}

Define \(P\) as the laser power, our target for control. \(P^*(l_v)\) is the optimal power at vector \(l_v\) that regulates the melt pool area \(A_c\) to some target \(A_{c,target}(l_v)\) and \(\nu(l_v)\) is the marking speed for that vector. 

\begin{definition}
    \underline{Vector} \(l_v\), beginning at timestep \(l_{v,0}\) and lasting for \(N_v\) timesteps, is a subdivision of a line the laser takes as it marks the part. The vector is partitioned according to the expected local thermal conditions, as presented in \cite{druzgalskiProcessOptimizationComplex2020}. Simply put, vectors are divided when transitioning between regions with differing amounts of solid material underneath. An example of the subdivision process can be seen in Figure~\ref{fig:overaching_controller_design}.

\end{definition}

Denote \(x,y,z\) as spatial coordinates in each direction, \(t\) as time, and \(h_s\) as the hatch spacing. We discretize the part into \(n\) voxels, hereafter referred to as \underline{elements}, with dimensions \(\Delta x\), \(\Delta y\), and \(\Delta z\). We let \(i,j,k\) index the \(x,y\), and \(z\) coordinates of the elements, respectively, choose \(\Delta x=\Delta y=h_s\) to capture the vector-level thermal fields, and \(\Delta z\) such that \(k\) becomes a layer index. This ensures each vector approximately scans over its own set of elements and that \(k-1\) is one layer down. Let \(l\) be an index for time discretized by \(\Delta t\) such that \(t=l\Delta t\). Matrices and vectors are represented via boldface symbols, and scalars are represented by italicized symbols.

\begin{definition}
Let $T(i,j,k,l)$ be the temperature at spatial index $(i,j,k)$ and time index $l$. Then, the \underline{temperature field} $\mathbf{T}(l)\in\mathbb{R}^n=[T(i_1,j_1,k_1,l),\dots,T(i_n,j_n,k_n,l)]$ is the temperature at all spatial indices at time $l$. We denote \(\mathbf{T}\) at vector \(l_v\) as \(\mathbf{T}(l_v)=\mathbf{T}(l_{v,0}+N_v)\), the temperature after all \(N_v\) timesteps of the vector.
 
\end{definition}

\begin{definition}
    The \underline{subsurface temperature}, denoted as \(T_b\), is the average temperature underneath a vector crossing \(N\) sets of spatial indices \(i_{sv},j_{sc},k_{sv}\), see Subsurface Temperature Prediction in Figure~\ref{fig:overaching_controller_design}. Crucially, for vector $l_v+1$, we calculate $T_b(l_v+1)$ from the temperature field \underline{just before scanning}, i.e. $\mathbf{T}(l_v)$, to capture the subsurface temperature during the scan of vector $(l_v+1)$. It relates to \(\mathbf{T}(l)\) via Eq.~\eqref{eq:subsurface_temp_defn}
    \begin{equation}\label{eq:subsurface_temp_defn}
        T_b(l_{v}+1)=\frac{1}{N}\sum_{i_{sv},j_{sv}k_{sv}} T(i,j,k-1,l_{v})
    \end{equation}

\end{definition}

\subsection{Vector-Level Feedforward Control Framework}\label{sec:FF_framework}

As shown in Figure~\ref{fig:overaching_controller_design}, the optimal power is computed per-vector to control the melt pool area. This is achieved by predicting how the part's underlying temperature, \(T_b(l_v+1)\), will affect the melt pool for each vector, and then adjusting the laser power accordingly. The underlying temperature is then updated using an FDM thermal model. The approach proceeds as follows:

\textbf{Step 1:} After vector \(l_v-1\), determine \(T_b(l_v)\) using Eq.~\eqref{eq:subsurface_temp_defn}. 

\textbf{Step 2:} Given \(T_b(l_v)\), scanning speed \(\nu(l_v)\), and target \(A_{c,target}\), perform the optimization over laser power, \(P\), in Eq.~\eqref{eq:FF_Control} to determine \(P^*(l_v)\).
\begin{equation}\label{eq:FF_Control}
\begin{split}
    &\text{Minimize:}\quad \bigl(A_c(P,\nu(l_v),T_b(l_v)) - A_{c,target}(l_v)\bigr)^2\\
    &\text{Subject To:}\quad P_{min} < P < P_{max},
\end{split}
\end{equation}
where \(A_c\) is obtained from the analytical melt pool model in Section~\ref{sec:analytical_model}, and \(P_{min}\) and \(P_{max}\) enforce machine limits. The choice of melt pool model simplifies this to a straightforward, analytical 1D optimization problem, making the optimization straightforward. 

\textbf{Step 3:} Use the optimized power \(P^*(l_v)\) in the FDM thermal model to predict \(\mathbf{T}(l_v)\), that is, simulate the model to the end of the vector. This ensures the predicted temperature field takes into account the optimized power.

\textbf{Step 4:} Determine \(T_b(l_v+1)\) from \(\mathbf{T}(l_{v})\) and repeat the process for all vectors in the part. 

\subsection{Coupled Analytical-Numerical Model}

Central to the proposed controller is the two-model structure, which is formulated in this work as a coupled analytical-numerical model. An analytical melt pool model enables fast and accurate optimization. However, it requires subsurface temperature measurements. These are achieved using a simple, part-scale FDM model to generate predictions of the thermal field at the vector level.

\subsubsection{Analytical Melt Pool Model}\label{sec:analytical_model}

This study bases the analytical melt pool model on the Rosenthal Equation, as formulated for a moving point heat source on a semi-infinite plate by \cite{rosenthalMathematicalTheoryHeat1941} and adapted by \cite{tangPredictionLackoffusionPorosity2017} for melt pool dimensions. Despite the existence of more accurate Gaussian beam models, the analytical solutions simplify the 1D optimization in Section~\ref{sec:FF_framework}. For the melt pool width \(W\) and length \(L\), these reduce to:

\begin{equation}\label{rosenthal_width}
    W\approx c_1\sqrt{\frac{P}{(T_m-T_b)v}}
\end{equation}
\begin{equation}\label{rosenthal_length}
    L\approx c_2\frac{P}{(T_m-T_b)}
\end{equation}
where \(c_1\) and \(c_2\) are fitting constants that depend on material properties and capture the complex physics involved in laser-material interaction. \(T_m\) is the melting point of the material, and \(T_b\) is the subsurface temperature before the vector is scanned.

\begin{remark}\label{remark:baseplateTemp}
    Note that in the case of scanning a single vector on a semi-infinite plate, \(T_b\) reduces to the base plate temperature.
\end{remark}

Fitting \(c_1\) and \(c_2\) ensures that complex melt pool dynamics are captured in the analytical model and do not need to be accounted for in the FDM thermal model. To ensure the effects of varying \(T_b\), such as overhang or turnaround regions, are captured, the baseplate temperature is varied while performing this calibration. Section~\ref{sec:analytical_model_calibration} explains this in detail.

\begin{figure}[h!]
    \centering
    \includegraphics[width=0.4\linewidth]{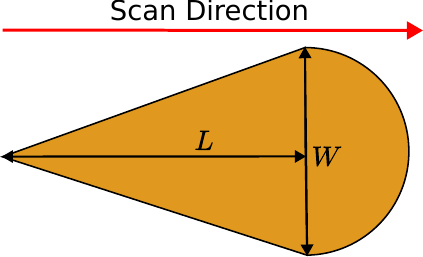}
    \caption{The geometric shape of the melt pool. A triangle of length  \(L\) and width \(W\) plus a half circle of diameter \(W\). The laser is scanning to the right.}
    \label{fig:meltpoolGeometry}
\end{figure}

The melt pool area can be derived using a geometric approximation where the melt pool is represented by a half circle of diameter \(W\) attached to a triangle of length \(L\), see Figure~\ref{fig:meltpoolGeometry}, giving the following equation for area.
\begin{equation}\label{rosenthal_area}
\begin{split}
        A_c(P,\nu,T_b)&=\frac{1}{2}WL+\frac{\pi}{8}W^2\\
        &=\frac{c_1c_2}{2}\sqrt{\frac{P^3}{(T_m-T_b)^3 \nu}}+\frac{\pi c_1^2}{8
        }\frac{P}{(T_m-T_b)\nu}
\end{split}
\end{equation}

This results in a scale-agnostic and explicit analytical model for melt pool area, enabling fast optimization over \(P\) to regulate melt pool area. However, an accurate prediction of \(T_b\) is needed. 

\subsubsection{Finite Difference Method Thermal Model}\label{sec:thermal_model}

To predict \(\mathbf{T}\) and thus, \(T_b\), we derive a simplified conduction-only FDM model. Thus, radiative heat transfer, Marangoni convection, latent heat, and other effects are ignored. While important at the microscale, these effects on \(A_c(P,\nu,T_b)\) are captured in the calibration of the melt pool model (\(c_1\) and \(c_2\)). Temperature dissipation from these effects is approximated by tuning the intensity of a volumetric heat input, described later in Eq.~\eqref{eq:volumetric_heat_source}, using the process in Section~\ref{sec:thermal_model_calibration}. This significantly reduces computational complexity, enabling faster simulation. Additionally, the entire part is treated as solid material, and, as powder has orders of magnitude less thermal conductivity \cite{weiThermalConductivityMetal2018}, it can be treated as insulation over the short time the laser is scanning the surface. As such, the top of the part is treated as a convection boundary condition, the sides are treated as insulation, and the bottom is treated as a constant temperature boundary condition as it is adhered to the build plate.

Similar to the process in \cite{heGeneralizedSmartScanIntelligent2024}, the governing equation for heat conduction with thermal diffusivity \(\alpha\) and conductivity \(k_{cond}\) is:
\begin{equation}\label{heat_conduction_govern}
    \frac{\partial^2 T}{\partial x^2}+\frac{\partial^2 T}{\partial y^2}+\frac{\partial^2 T}{\partial z^2}+\frac{Q_{laser}}{k_{cond}}=\frac{1}{\alpha}\frac{\partial T}{\partial t}
\end{equation}
where \(Q_{laser}\) is the volumetric heat input.

This paper uses a hemispherical Goldak heat source as described in \cite{goldakNewFiniteElement1984} for \(Q_{laser}\), which can be found below:
\begin{equation}\label{eq:volumetric_heat_source}
    Q_{laser}(x,y,z)=f\frac{6\sqrt{3}\eta P}{r_xr_yr_z\pi\sqrt{\pi}}e^{-\frac{3(x-x_c)}{r_x^2}-\frac{3(y-y_c)}{r_y^2}-\frac{3(z-z_{max})}{r_z^2}}
\end{equation}
where \(f\) is a tuning factor applied to the heat input, \(\eta\) is the absorptivity of the material, \(r_x,r_y,\) and \(r_z\) are the beam radius in the \(x,y,\) and \(z\) directions respectively. For convenience, we assume \(r_x=r_y=r_z\) and set the beam radius to half the spot size. \(x_c\) and \(y_c\) are the location within the part where the laser is being applied, and \(z_{max}\) is the top surface.

The FDM and the forward Euler method can then be used to discretize Eq.~\eqref{heat_conduction_govern} into:

\begin{equation}\label{heat_conduction_FDM}
\begin{split}
    &\frac{T(i+1,j,k,l)-2T(i,j,k,l)+T(i-1,j,k,l)}{\Delta x^2}\\
    &+\frac{T(i,j+1,k,l)-2T(i,j,k,l)+T(i,j-1,k,l)}{\Delta y^2}\\
    &+\frac{T(i,j,k+1,l)-2T(i,j,k,l)+T(i,j,k-1,l)}{\Delta z^2}\\
    &+\frac{Q_{laser}(i,j,k,l)}{k_{cond}}=\frac{1}{\alpha}\frac{T(i,j,k,l+1)-T(i,j,k,l)}{\Delta t}
\end{split}
\end{equation}

As such, \(T(i,j,k,l)\) gives the temperature of element \(i,j,k\) at time \(t=l\Delta t\). The elements on the boundaries have corresponding discretizations to maintain boundary conditions, which can be found in \cite{heGeneralizedSmartScanIntelligent2024}. These boundary terms have constants, not dependent on \(\mathbf{T}\), which are collected into \(\mathbf{Q_{disturbance}}\). 

\(Q_{laser}\) is integrated over space for each element and then divided by the element's volume. This is needed since, for arbitrary scan patterns, we cannot guarantee that the centroid of each element will align with the laser location. The averaging guarantees heat input is conserved regardless of scan pattern. Assuming the elements are \(\Delta x\), \(\Delta y\), and \(\Delta z\) in each direction and discretizing in time, we get the following:
\begin{equation}\label{eq:integral_heat_source}
\begin{split}
    &{Q_{laser}}(i,j,k,l)=f\frac{\Delta t}{\alpha}\frac{\eta P(l)}{4 \Delta x\Delta y \Delta z}\\
    &\left[\erf{\left(\frac{\sqrt{3}}{r_x}(x_c(l)+\frac{\Delta x}{2}-X(i))\right)}-\erf{\left(\frac{\sqrt{3}}{r_x} (x_c(l)-\frac{\Delta x}{2}-X(i))\right)}\right]\\
    &\left[\erf{\left(\frac{\sqrt{3}}{r_y}(y_c(l)+\frac{\Delta y}{2}-Y(j))\right)}-\erf{\left(\frac{\sqrt{3}}{r_y} (y_c(l)-\frac{\Delta y}{2}-Y(j))\right)}\right]\\
    &\left[\erf{\left(\frac{\sqrt{3}}{r_z}(z_{max}+\frac{\Delta z}{2}-Z(k))\right)}-\erf{\left(\frac{\sqrt{3}}{r_z} (z_{max}-\frac{\Delta z}{2}-Z(k))\right)}\right]
\end{split}
\end{equation}

\noindent where \(X(i),Y(j),\) and \(Z(k)\) denote the centroid of the element \((i,j,k)\) in space, \(P(l)\) denotes the laser power at timestep \(l\), \(\Delta t\) is the timestep length, and \(x_c(l),y_c(l),\) and \(z_{max}\) denote the laser position at each timestep.

\(Q_{laser}(i,j,k,l)\) can be computed for all elements, combined into a vector \(\mathbf{Q_{laser}}(l)\) and added to \(\mathbf{Q_{disturbance}}\) to give the total input to the system.
\begin{equation}\label{convection}
    \mathbf{u}(l)=\mathbf{Q_{laser}}(l)+\mathbf{Q_{disturbance}}
\end{equation}

Rewriting Eq.~\eqref{heat_conduction_FDM} in matrix form yields a system of linear equations well-suited for state-space representation, which enables efficient stepping through time.
\begin{equation}\label{heat_conduction_state_space}
    \mathbf{T}(l+1)=\mathbf{A}\mathbf{T}(l)+\mathbf{B}\mathbf{u}(l)
\end{equation}
where \(\mathbf{A}\) is the state matrix, \(\mathbf{B}\), the input matrix, is the identity matrix, and \(\mathbf{u}(l)\) is the input plus the disturbance vector from the boundary conditions at timestep \(l\). 

Finally, timesteps are mapped onto the scan path, and the model is lumped to the vector level. Unlike in \cite{heGeneralizedSmartScanIntelligent2024}, the time in-between vectors is accounted for and treated as a vector with zero power. This includes the skywriting time, that is, when the laser maintains constant velocity along the vector and only accelerates and decelerates outside the part.
\begin{equation}\label{lumped_model}
\begin{split}
    &\mathbf{T}(l_v)=\mathbf{T}(l_{v,0}+N_v)=\mathbf{A}_v\mathbf{T}(l_{v,0})+\mathbf{b}_v\\
    &A_v\equiv A^{N_v};\;\mathbf{b}_v \equiv \sum_{m=0}^{N_v-1}\mathbf{A}^{N_v-1-m} \mathbf{B}\mathbf{u}(m)
\end{split}
\end{equation}

Once \(P^*(l_v)\) is used to simulate \(\mathbf{T}\), \(T_b(l_v+1)\), as is required for the analytical melt pool model in Section~\ref{sec:analytical_model}, is extracted. We rewrite Eq.~\ref{eq:subsurface_temp_defn} as a time-varying output function to fit into the linear system framework.
\begin{equation}\label{eq:Tb_scope}
    T_b(l_v+1)=\mathbf{C}(l_v+1)\mathbf{T}(l_{v})
\end{equation}
where \(\mathbf{C}(l_v+1)\) is the matrix form of the summation in Eq.~\eqref{eq:subsurface_temp_defn}. This enables efficient measurement of \(T_b(l_v+1)\) before scanning the vector, keeping the part-scale FDM thermal model out of the optimization step.

\begin{figure}[h!]
    \centering
    \includegraphics[width=0.5\linewidth]{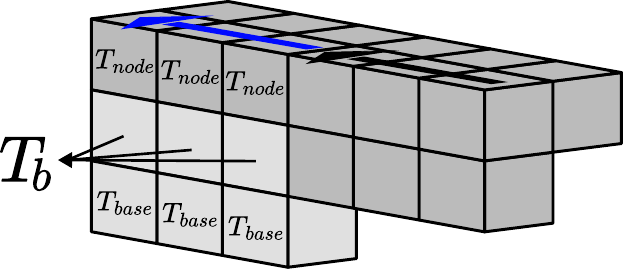}
    \caption{In overhang regions, where only one layer of part exists, \(T_b\) is determined by computing the heat transfer through the powder using an analytical solution. Two layers of powder are added underneath the part, with the bottom being set at the baseplate temperature \(T_{base}\) and the top of the analytical solution at the node temperature \(T_{node}\).}
    \label{fig:subsurfaceTemp_pic_overhang}
\end{figure}

When scanning over a region that is a single layer thick, such as overhang regions, for the first time, the previous layer is undefined, and so \(T_b\) is computed via 1D heat transfer between the current node being scanned and the powder two layers below. As shown in Figure~\ref{fig:subsurfaceTemp_pic_overhang}, the midpoint, one layer down, is where \(T_b\) is measured, resulting in Equation \eqref{eq:Tb_scope_overhang}. Powder is assumed to have a density \(48\%\) of that of the nominal material \(\rho_{powder}=0.48 \rho\) \cite{denlingerThermalModelingInconel2016} and a conductivity one tenth \(k_{powder}=0.1k_{cond}\): 
\begin{equation}\label{eq:Tb_scope_overhang}
\begin{split}
    &T_b(l_v+1)=T_{node}(l_{v,0})+\frac{4(T_{base}-T_{node}(l_{v,0}))}{\pi}\\
    &\cdot \sum_{m=0}^{50}\left(\frac{(-1)^m}{2m+1}\right) \exp\left(-\left(\frac{\pi(2m+1)}{4\Delta z}\right)^2 \alpha_{powder}\Delta \tau \right)\cdot \cos\left(\frac{\pi}{4}(2m+1)\right)
\end{split}
\end{equation}
where \(T_{node}\) is the temperature of the node before being scanned, \(T_{base}\) is the baseplate temperature, \(\alpha_{powder}\) is the diffusivity of the powder, and \(\Delta \tau\) is the time between predicting the subsurface temperature and when the node will be scanned.

The number of layers included in the model is restricted to reduce the computational burden, similar to \cite{heGeneralizedSmartScanIntelligent2024}. However, we set the bottom layer to a constant temperature field equal to the temperature field in that layer before the scanning starts to enable more accurate absolute temperature measurements. Following a similar process as in \cite{heGeneralizedSmartScanIntelligent2024}, it was determined that \(P^*(l_v)\) stopped varying on the geometry given in Figure~\ref{fig:steppedPyramid_scan_pattern} only after 30 layers were added to the simulation. As such, 30 layers were chosen to represent the part during scanning. 
 
At a resolution sufficient to resolve vector-level thermal fields, the timestep required to keep the model stable is too small to model the recoating and interlayer dwell explicitly. As such, the dwell time is modeled by propagating heat down in Z using 1D analytical heat transfer and then spreading heat in X and Y via Gaussian blurring. A full description of the process has been relegated to \ref{sec:interlayer_dwell}.

As such, a lightweight part-scale thermal model, coupled with an analytical melt pool model, has been created to predict melt pool dimensions throughout a part. This works at the part scale on a per-vector basis without requiring a high-resolution grid of elements. Critically, the two-model structure is modular: while we chose an FDM thermal model and a Rosenthal-based melt pool model, any sufficiently fast and accurate thermal model can be substituted to provide subsurface temperature data, and any reduced-order melt pool model can be used to link temperature data to melt pool geometry. By decoupling the melt pool dynamics from the part-scale thermal model, the part-scale thermal model moves outside the optimization, enabling scale-agnostic control. Similarly, the optimization target could be changed to any feature that can be tied to \(\mathbf{T}\). In our case, key assumptions and simplifications in the model necessitate careful calibration outlined in the next section. 

\section{Model Calibration}\label{text_modelCalibration}

The model calibration is broken down into two main steps. First, the melt pool model is calibrated using single-track scans on a heated baseplate to find \(c_1\) and \(c_2\). Second, the thermal model is calibrated by tuning \(f\) to minimize Eq.~\eqref{eq:ffactor_tuning} on a custom geometry.

\subsection{Experimental Setup} \label{text:experimentalSetup}

Experiments were conducted using an open-architecture PANDA 11 LPBF machine, which offers vector-level control over process parameters. The system features a 500W IPG Photonics \(1070\,nm\) fiber laser with a spot size (D4Sigma) of \(78\,\mu m\) as measured during installation, paired with a SCANLABS hurrySCAN galvo scanner plus SCANLABS varioSCAN system for beam control. Vectorwise scheduling of power was achieved using an open-source XML format \cite{OASISChallengeBaseline2022}.

In situ monitoring is achieved using two coaxial sensors. A high-resolution camera (19.5\,\(\mu m\)/pixel, 2\,kHz sample rate, 700–1000\, nm bandwidth) and a photodiode sensor (50\,kHz sample rate, 700–1000\, nm bandwidth) capturing real-time melt pool dynamics. The camera trigger, galvo positions, and photodiode signals are all recorded via a SCANLABS Open Interface Extension card, enabling each signal to be tied to a position. An Optris PI 640 G7 IR camera with a resolution of \(0.64\,mm/\text{pixel}\) was used to image the entire build plate at \(125\,Hz\) during the print. Additionally, an induction-heated bed enables control of the baseplate temperature, while a crossflow of nitrogen gas (\(3.26\,m/s\)) minimizes the vapor plume-laser interaction, ensuring a stable processing environment.

For calibrating both the melt pool model and the FDM thermal model, thin \(146\, mm \times 146\, mm\) plates, \(3.1\,mm\) thick for IN718 and \(0.8\,mm\) thick for 316LSS, were bolted onto the baseplate. The material properties used for the FDM thermal model can be found in Table \ref{table:material_properties}.

\begin{table}%
\caption{Simulation Material Properties}\label{table:material_properties}
\begin{tabular*}{\tblwidth}{@{}LLL@{}}
\toprule
  Property & IN718 & 316LSS\\ %
\midrule
Density (\(kg\,m^{-3}\)) & 8260 & 7900\\
Heat Capacity (\(J\,kg^{-1}\,K^{-1}\)) & 543 & 434\\
Thermal Conductivity (\(W\,m^{-1}\,K^{-1}\)) & 14.90 & 13.96\\
Convection Coefficient (\(W\,m^{-2}\,K^{-1}\)) & 20 & 20\\
Melting Temperature (\(K\)) & 1610 & 1710\\
Ambient Temperature (\(K\)) & 293 & 293\\
Absorptivity & 0.33 & 0.33\\
\bottomrule
\end{tabular*}
\end{table}

\subsection{Calibration of the Analytical Model }\label{sec:analytical_model_calibration}

\begin{figure}[h!]
    \centering
    \includegraphics[width=0.425\linewidth]{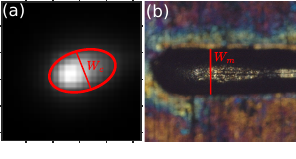}
    \caption{(a) An image of the melt pool through the coaxial camera with the identified ellipse drawn. \(W_c=151\,\mu m\). (b) The same single-track via optical microscope. \(W_m=149\,\mu m\)}
    \label{fig:meltpool_image}
\end{figure}

A series of single-track scans at different levels of $P$, $v$, and $T_b$ were performed on the aforementioned plates to calibrate the analytical melt pool model. The values of $T_b$ were varied using the base plate's induction heater to mimic the elevated temperatures encountered in overhangs and other heat islands of real printed geometries. This ensures that the calibration is valid not only for single tracks but also for full geometries. Each scan track was \(20\, mm\) in length, with a delay of \(500\, ms\) between successive vectors and a minimum spacing of \(2\, mm\) to avoid interference between scans. 

\(P\), \(\nu\), and \(T_b\), seen in Table \ref{table:analytical_calibration_parameters}, were chosen to span typical processing windows for Inconel 718 and 316L Stainless Steel in full LPBF builds. \(433\,^\circ C\) was the highest the buildplate would reach. Each scan line was repeated twice to ensure sufficient images to capture variations in the dimensions. As such, a total of 720 single-track scans were performed per material to calibrate the analytical model. The coaxial camera was calibrated to identify the gray value closest to the solid-liquid transition, and melt pool dimensions were determined by fitting an ellipse to the pixels brighter than that value. As shown in Figure~\ref{fig:meltpool_image}, a subset of the measured widths was validated using optical microscope images to confirm the accuracy of the measurements.

\begin{table}%
\caption{Parameter sweep values used for single track scans. Tracks were fused using all possible combinations of these parameters.} 

\label{table:analytical_calibration_parameters}
\begin{tabular*}{\tblwidth}{@{}ll@{}}
\toprule
Parameter & Sweep (Start:Increment:End) \\ 
\midrule
Power \(P\) (\(W\))                & 100\,:\,40\,:\,420 \\[2pt]
Speed \(\nu\) (\(mm\, s^{-1}\))             & 500\,:\,150\,:\,1850 \\[2pt]
Subsurface Temp. \(T_b\) (\(^\circ C\)) & 50, 200, 300, 433 \\
\bottomrule
\end{tabular*}
\end{table}
\subsection{Calibration of the FDM Thermal Model }\label{sec:thermal_model_calibration}

Once the analytical melt pool model is calibrated, the FDM thermal model can be calibrated. This requires modifying the optimal value of the tuning factor \(f\) in the volumetric heat source, Eq.~\eqref{eq:integral_heat_source}, until the resulting \(P^*\) minimizes the variance of the melt pool area. This process is iterative and requires accurate measurements. As such, this process is done in 2D on thin plates in the same manner as Section~\ref{sec:analytical_model_calibration}.

\begin{figure}
    \centering
    \includegraphics[width=0.6\linewidth]{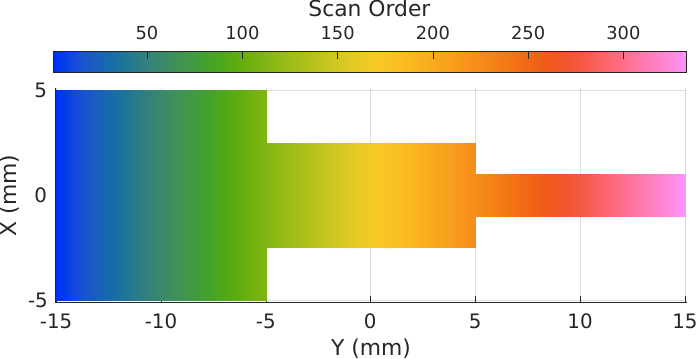}
    \caption{The scan pattern for the "stepped pyramid" used to calibrate the FDM thermal model. The laser scans up and down (x-dir) from left to right (y-dir). There are 333 vectors in the scan, with 111 for each width. Each width has a characteristic subsurface temperature that the FDM thermal model tries to capture.}
    \label{fig:steppedPyramid_scan_pattern}
\end{figure}

We chose the 2D geometry depicted in Figure~\ref{fig:steppedPyramid_scan_pattern}, hereafter referred to as the "stepped pyramid geometry", to perform this calibration on. This geometry contains three distinct vector lengths, which provide three distinct \(T_b\) values, thus three \(A_c\) values (see Figure~\ref{fig:all_calibration}b), that the FDM thermal model aims to capture. In simulation, the plate is modeled as a \(40\, mm \times 40\, mm\) region, \(1.2\,mm\) or 30 layers deep, and the laser follows the scan pattern shown in Figure~\ref{fig:steppedPyramid_scan_pattern}. 

\begin{assumption}\label{assump:nominal_parameters}
    The nominal scan parameters, found in Table \ref{tbl:nominal_scan_parameters}, are sufficient to minimize defects in bulk material (such as a simple cube).
\end{assumption}

As is common in controls literature, we selected a setpoint \(P_{nominal}\) that was close to the desired behavior, see Assumption \ref{assump:nominal_parameters}. As such, the calibration should result in a \(P^*\) close to \(P_{nominal}\) and with a corresponding \(A_{c,nominal}\) under steady-state conditions representing printing bulk material. In regions with more complex thermal responses, the controller should choose a \(P^*\) such that \(A_c\) matches \(A_{c,nominal}\), that is, we set \(A_{c,target}=A_{c,nominal}\).

With this in mind, two key parameters must be determined \(f\) and \(A_{c,nominal}\). This is done via the following process.

\textbf{Step 1:} Pick a value for \(f\), this can either be based on prior knowledge or simply random. Reasonable values are between 0.5 and 5.

\textbf{Step 2:} Tune \(A_{c,target}\) until \(P^*(l_v)\approx P_{nominal}\) in the region \(X\in(-10\,mm,-5\,mm)\) on the stepped pyramid geometry in Figure~\ref{fig:steppedPyramid_scan_pattern}. This excludes initial transient effects and does three things. It determines \(A_{c,nominal}\), ensures that in bulk material the chosen power is close to the nominal value, and compensates for modeling errors that cause a discrepancy between the predicted \(A_c\) and the experimentally measured \(A_c\). By doing this all at once, we partially compensate for both the higher-order terms not captured in the analytical melt pool model by \(c_1\) and \(c_2\), as well as the limited physics in the FDM thermal model.

\textbf{Step 3:} Simulate \(P^*\) for the entire stepped pyramid geometry and then apply that power sequence in an experimental scan of Figure~\ref{fig:steppedPyramid_scan_pattern}. In the same manner as in Section~\ref{sec:analytical_model_calibration}, the melt pool area is measured using the coaxial camera for each vector. 

\textbf{Step 4:} To quantify the melt pool regulation, a normalized error metric is defined:
\begin{equation}\label{eq:ffactor_tuning}
    \epsilon(f)=\frac{\|\mathbf{A}_c-\bar{A}_c\|_2}{\bar{A}_c},
\end{equation}
where \(\mathbf{A}_c\) is the vector of measured melt pool areas across the part and \(\bar{A}_c\) is the corresponding average melt pool area. Using Eq.~\eqref{eq:ffactor_tuning}, compute the normalized error, \(\epsilon(f)\).

\textbf{Step 5:} Pick a new \(f\) and repeat the process until \(\epsilon(f)\) is minimized.

This iterative process adjusts the FDM thermal model calibration, \(f\) and \(A_{c,target}\), until the optimal power sequence \(P^*\) minimizes the variance of the melt pool area. This calibration assumes that we can extend the 2D calibration to complex 3D geometries, as will be validated in the experiments reported in Section~\ref{sec:validation_test_setup}.

\subsection{Calibration Results}

\begin{figure}
  \centering
  \includegraphics[width=0.9\linewidth]{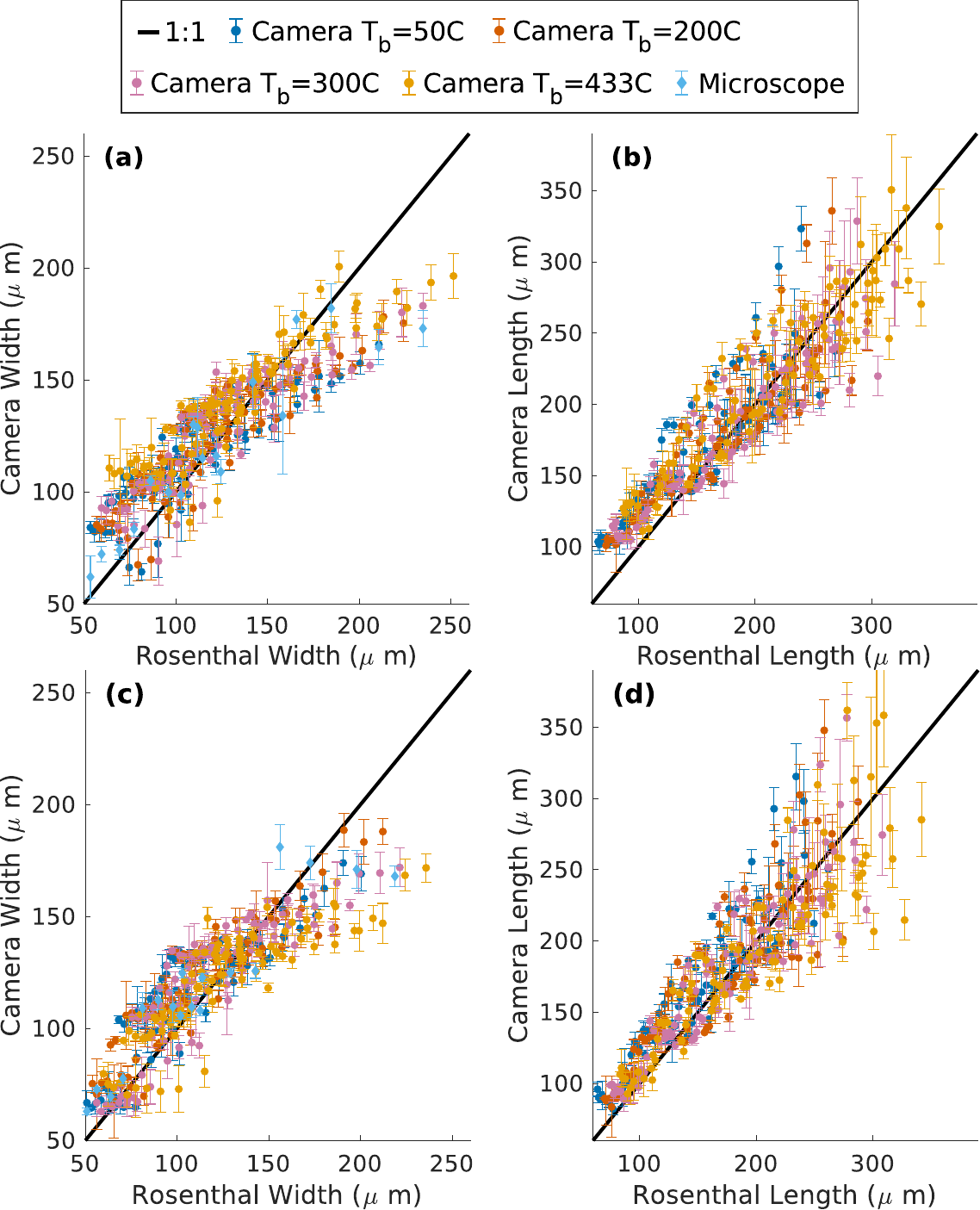}
  \caption{Measured versus Rosenthal-model–predicted steady-state single-track melt pool widths in IN718 (a), in 316LSS (c), and lengths in IN718 (b), in 316LSS (d). The baseplate temperature, \(T_b\), is varied for four separate values, and the widths and lengths are measured using a coaxial camera. Additional measurements are made of the width via an optical microscope.}
  \label{fig:rosenthal_width_length}
\end{figure}

Using the measured melt pool widths and lengths, Eqs. \eqref{rosenthal_width} and \eqref{rosenthal_length} were fit by least squares for both IN718 and 316LSS. Figure~\ref{fig:rosenthal_width_length} illustrates these fits, and the resulting parameters and \(R^2\) values can be found in Table \ref{table:rosenthal_fits}. While both models capture the overall trends, the width fits exhibit systematic deviations and overall low \(R^2\) values. These discrepancies primarily arise from the Rosenthal model's assumption of a point heat source; in reality, the laser beam follows a Gaussian distribution, skewing the fit. Incorporating a Gaussian-based model could reduce these errors, but it would also eliminate the availability of an explicit equation for dimensions and thus substantially increase the computational cost during optimization. Despite these systematic deviations and low \(R^2\) values, the model captures overall trends sufficiently for feedforward control, as the optimization relies on relative changes, not absolute ones.

\begin{figure*}
    \centering
    \includegraphics[width=0.98\linewidth]{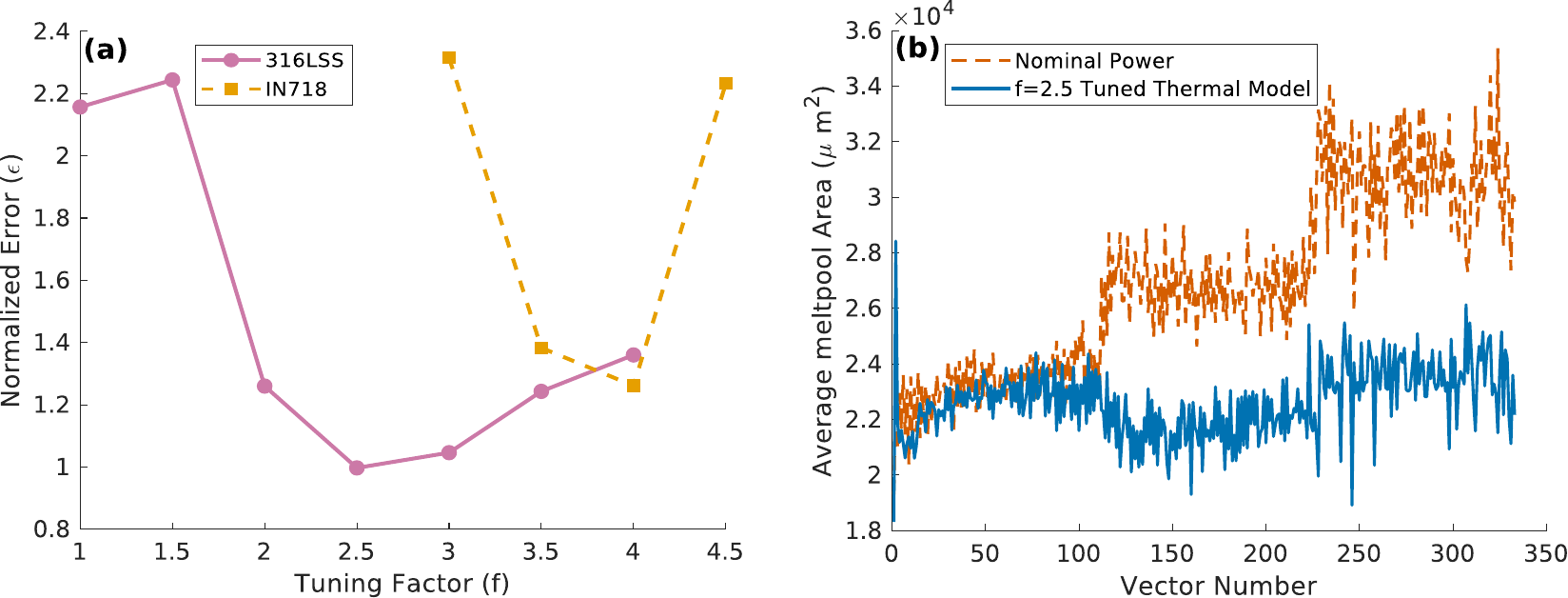}
    \caption{(a) Eq.~\eqref{eq:ffactor_tuning} plotted against \(f\). By tuning the heat input to the FDM thermal model in Eq.~\eqref{eq:integral_heat_source}, the model can be calibrated to predict subsurface temperatures accurately. (b) Averaged melt pool area for the stepped pyramid geometry over three tests on 316LSS. The FDM thermal model tuned with \(f=2.5\) shows a 54\% drop in normalized error of the melt pool area compared with the nominal case.}
    \label{fig:all_calibration}
\end{figure*}

For the calibration of the FDM thermal model, the optimal tuning factor was determined to be \(f=4\) for IN718 and \(f=2.5\) for 316LSS. For both materials \(A_{c,target}=0.0164\,mm^2\) resulted in \(P^*(l_v)\approx P_{nominal}\) in the bulk region. Other than power, the scanning parameters match the nominal parameters shown in Table \ref{tbl:nominal_scan_parameters}. Figure~\ref{fig:all_calibration}a shows the normalized error \(\epsilon(f)\) as a function of \(f\), illustrating how iterating \(f\) minimizes the error. Once minimized, we expect the controller to reduce melt pool area variance compared to the nominal power, which can be seen in Figure~\ref{fig:all_calibration}b for 316LSS. This shows that the tuning factor of \(f=2.5\) decreases the melt pool variation, thus confirming that the FDM thermal model captures the characteristic \(T_b\) value in those three regions. 

Overall, under \(3000\) vectors per material were scanned on 2D plates to calibrate both the analytical melt pool model and the FDM thermal model. This is orders of magnitude less than contemporary data-driven works such as \cite{lapointePhotodiodebasedMachineLearning2022}, enabling this calibration to be performed in a simpler manner, without requiring consumables like powder, and in less time.

\begin{table}
  \caption{Rosenthal-model fit constants and coefficients of determination corresponding to Fig.~\ref{fig:rosenthal_width_length}.}
  \label{table:rosenthal_fits}
  \begin{tabular*}{\tblwidth}{@{}llllc@{}}
    \toprule
    Material & Fit constant & $R^{2}$ \\
    \midrule
    IN718    & $c_{1}=261$ & 0.43 \\
             & $c_{2}=499$ & 0.74 \\
    316LSS   & $c_{1}=256$ & 0.45 \\
             & $c_{2}=529$ & 0.76 \\
    \bottomrule
  \end{tabular*}
\end{table}

\section{Validation Test Setup}\label{sec:validation_test_setup}

\begin{figure}
    \centering
    \includegraphics[width=0.5\linewidth]{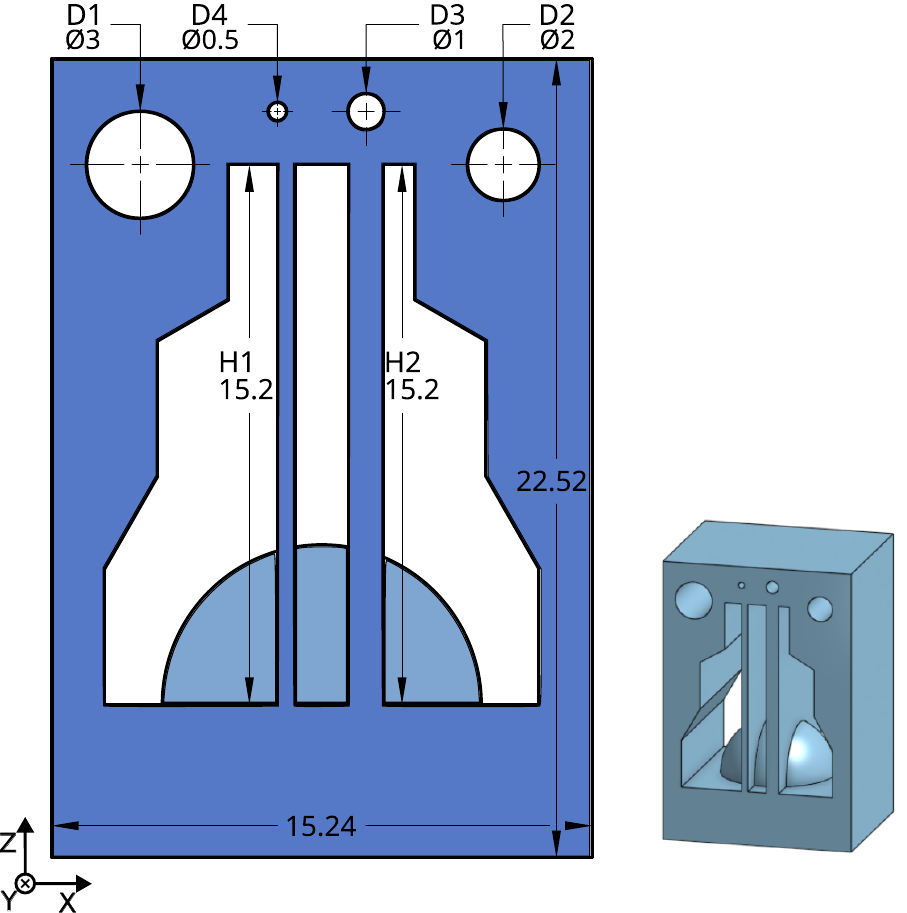}
    \caption{Key dimensions of the part from \cite{lapointePhotodiodebasedMachineLearning2022} used in this study. All units are in \(mm\). The depth is \(12.7\,mm\) into the page, and the \(9\,mm\) hemisphere is centered on the part. Key dimensions are labeled with \(D1\), \(D2\), \(D3\), \(D4\), \(H1\) and \(H2\). A full drawing of the part is available in the supplemental material.}
    \label{fig:cologne_bottle_dims}
\end{figure}

In this section, a full 3D geometry is used to validate that the calibration of both the analytical model on single tracks and the FDM thermal model in 2D extends beyond 2D into 3D, where overhangs and changing cross-sectional geometry provoke different changes in subsurface temperature. As such, the geometry shown in Figure~\ref{fig:cologne_bottle_dims}, hereafter referred to as the "cologne bottle geometry" and described in \cite{lapointePhotodiodebasedMachineLearning2022}, was chosen. Full drawings can be found in the supplemental material. It was chosen for its combination of thin walls, overhangs, and holes. The key dimensions of the part are shown in Figure~\ref{fig:cologne_bottle_dims}. \(A_{c,target}\) was kept constant at the value found in Section~\ref{sec:thermal_model_calibration} for the entire print. Nominal scan parameters and slicing parameters can be found in Table \ref{tbl:nominal_scan_parameters}, and, where possible, the scanning pattern was kept the same as in \cite{lapointePhotodiodebasedMachineLearning2022}. Scan vectors go across the part, skywrite for approximately \(1.8\,ms\), and turn back around in a snake pattern. Due to machine limitations, skywriting was turned off for regions with subdivided vectors in the controlled samples; Nominal samples always had full skywriting. This increases the melt pool area at turnarounds for the Feedforward (FF) Controlled samples due to the lack of time for the material to cool between vectors. 

\begin{figure}
    \centering
    \includegraphics[width=0.4\linewidth]{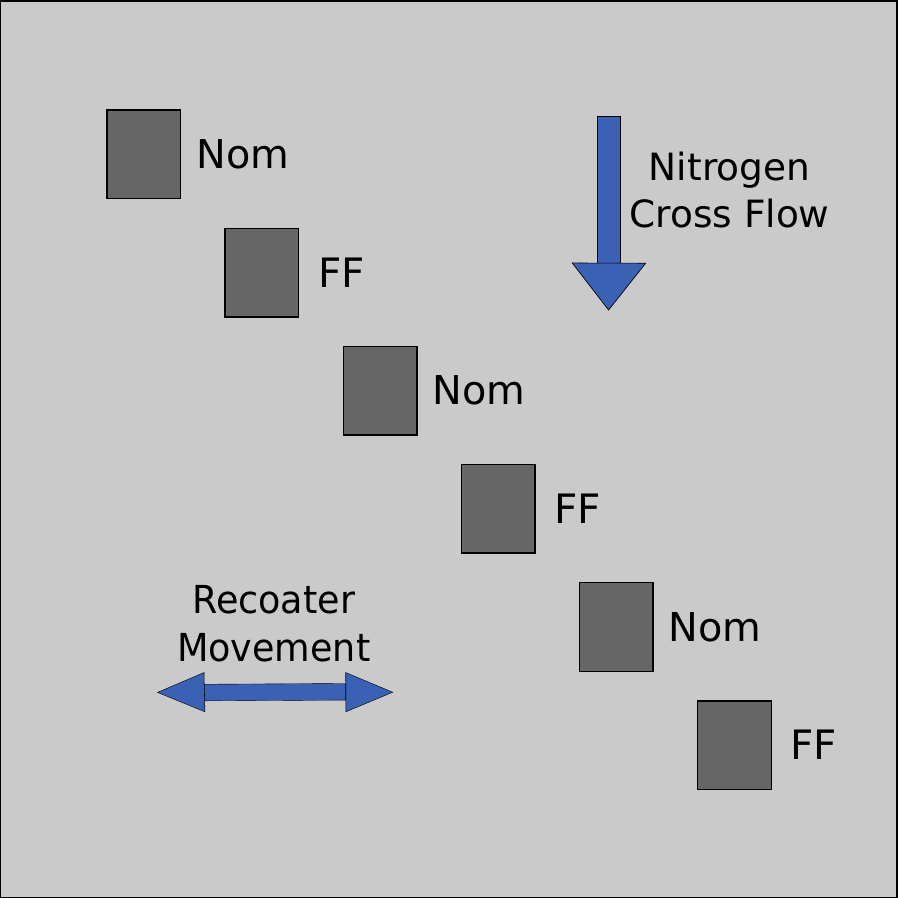}
    \caption{A schematic of how the samples were laid out on the buildplate with nitrogen crossflow and recoater direction labeled. The top left samples were chosen for further analysis.}
    \label{fig:sample_orientation}
\end{figure}

To test the model's ability to work with various materials, the part was printed in both IN718 and 316LSS. The print orientation and setup can be found in Figure~\ref{fig:sample_orientation}. Powder with specifications in Table \ref{table:powder_particles} from Carpenter Additive (Tanner, AL, USA) and the same machine setup as in Section~\ref{text:experimentalSetup} was used. 

\begin{table}[h]
\caption{Nominal scan and slicing parameters}
\label{tbl:nominal_scan_parameters}
\begin{tabular*}{\tblwidth}{@{}lcc@{}}
\toprule
Parameter & IN718 & 316LSS\\
\midrule
Power (\(W\)), $P_{nominal}$   & 220  & 290\\
Speed (\(mm/s\))  & 1000   & 1200\\
Layer Thickness (\(\mu m\))  & 40   & 40\\
Hatch Spacing (\(\mu m\))  & 90   & 90\\
Hatch Rotation (\(\deg\))  & 67   & 67\\
Hatch Offset (\(\mu m\))  & 90   & 90\\
\bottomrule
\end{tabular*}
\end{table}

The part was sliced, and the algorithm of Section~\ref{sec:FF_framework} was used to generate a power sequence for the cologne bottle. Three copies of Nominal and FF Controlled samples, see Figure~\ref{fig:sample_orientation}, were printed for each material. This ensured that sample-to-sample variance in the in situ data was taken into account. In situ photodiode data and galvo position were recorded as a proxy for the melt pool area, as the coaxial camera data were too noisy when powder was added. All 562 layers of the part were simulated in 5 hours using GPU-accelerated matrix computation on an RTX 4080, approximately the same time it took to print.

\begin{remark}\label{remark:photo_variation}
While photodiode signal is known to correlate with melt pool area \cite{clijstersSituQualityControl2014,fisherDeterminingMeltPool2018,craeghsFeedbackControlLayerwise2010}, it is essential to note that it does not measure the melt pool area directly. Our prior experiments suggest that the photodiode used in this work may yield a lower signal for thin-walled structures as compared to solid sections for the same melt pool area. 
\end{remark}

Two samples, one Nominal and one FF Controlled, from each material, were chosen for further analysis. These samples were cut in half and removed from the build plate using a wire EDM. For all samples, one half was X-ray CT scanned on a Bruker Skyscan 1273, with a focal spot size of \(4\,\mu m\), a \(1\,mm\) Molybdenum filter, at \(130\,kV\) and \(130\,\mu A\), and at a \(13.13\, \mu m/pixel\) resolution. Key dimensions listed on Figure~\ref{fig:cologne_bottle_dims} were compared to determine geometric accuracy. However, \(D4\) was too small to resolve in the resulting images.

Analysis of the porosity was conducted by grinding down and then polishing the other half up to a \(1\,\mu m\) diamond polish. The surface was then inspected using a Keyence VHX-S770E Microscope with a VHX-7020 Integrated Camera at a \(100\times\) objective for a resolution of \(1.01\,\mu m/\text{pixel}\). When comparing samples, both samples were imaged in the same pass to ensure consistent exposure. Pores were detected by performing an adaptive thresholding of the stitched image and then removing all pores that were only one pixel large.

\begin{table}[h]
\centering
\caption{Powder particle size distribution.}
\label{table:powder_particles}
\begin{tabular*}{\tblwidth}{@{}lccc@{}}
\toprule
Material               & D10 (\(\mu m\)) & D50 (\(\mu m\)) & D90 (\(\mu m\)) \\
\midrule
Inconel 718            & 19  & 31 & 50 \\
316L Stainless Steel   & 18   & 32   & 51   \\
\bottomrule
\end{tabular*}
\end{table}

\section{Results and Discussion}\label{text_results}

\subsection{Overall Layerwise Results}\label{sec:layerwise_results}

Full videos showing \(P^*\) and the resulting photodiode signal for one set of samples in both materials, across all layers, are available in the supplementary materials. Additionally, thermal camera data from key layers are provided for both materials. "IN718\_and\_316LSS\_Optimal\_Powers.mp4" reveals significant variation in \(P^*\) across the part, with notable decreases in overhang regions. "IN718\_and\_316LSS\_Photodiode\_Signal.mp4" demonstrates that the controller significantly reduces variation, and "IN718\_Thermal\_Camera.mp4" and "316LSS\_Thermal\_Camera.mp4" show that the controller significantly reduced overheating. In layers 488 and 489, the overhang region, the Nominal case cooled significantly slower than in the FF Controlled case, showing reduced overheating. Layer 150 illustrates the similarity between the two cases, where heat can easily dissipate. In contrast, layer 550 shows that the initial temperature increases with height; however, the FF Controlled sample is cooler due to the lower heat input.

Figure~\ref{fig:photodiodeVariation_and_power_layerwise} summarizes the results in the supplementary materials. It displays the average power in each layer and the mean variation of the photodiode for all three samples against the layers in the part. This was achieved by computing the standard deviation of the signal in each layer for both the FF Controlled and Nominal samples, and then plotting it on a per-layer basis. Both plots are smoothed with a 20-layer moving average.  

Regarding the power, both materials showed a similar trend, despite having different average powers. In the bottom regions, where the cross section is a rectangle, the power remained close to \(P_{nominal}\). This is to be expected as this represents bulk material where Assumption \ref{assump:nominal_parameters} holds. The controller reduced the power significantly as the part's cross-section changed from a solid rectangle to isolated pillars due to the shorter vectors in the pillars requiring less power for the same melt pool area. In regions where the pillars widened, the power was further reduced due to the overhang, reducing overheating. In the thicker areas of the pillar, the power was maintained at a fairly constant level. Finally, at the top of the part, the average power increased again due to returning to a solid rectangle; however, there was an increase in the latent part temperature, as shown by "IN718\_Thermal\_Camera.mp4" and "316LSS\_Thermal\_Camera.mp4" in the supplementary material, so the average power did not return to \(P_{nominal}\) and instead was lower to keep the melt pool size constant.

As a proxy for melt pool area variation, the photodiode variation was examined. The first 100 and the last few layers showed low variation, but the FF Controlled samples showed slightly worse variation. The cross-section of the part in these regions is a pure rectangle, and so the Nominal samples have very little variation across the part. In contrast, the FF Controlled samples' power modulation introduces slight changes to the photodiode signal, increasing the measured variation. However, layers 100-550, with constantly changing cross-sections, exhibited a significant increase in photodiode signal variation, which the controller partially mitigated. In this middle region, the FF controller exhibited an average decrease in photodiode variation per layer of \(5.3\%\) in IN718 and \(8.3\%\) in 316LSS for an overall decrease of \(6.8\%\), implying reduced melt pool variation throughout the build. However, it was unable to fully compensate for these variations, in part, due to the controller in this work only dividing vectors in overhang regions, thus being unable to compensate for the spike in melt pool area at the beginning of a vector. Additionally, as described in Remark \ref{remark:photo_variation}, the photodiode is not a perfect proxy for melt pool area.

\begin{figure*}
    \centering
    \includegraphics[width=\linewidth]{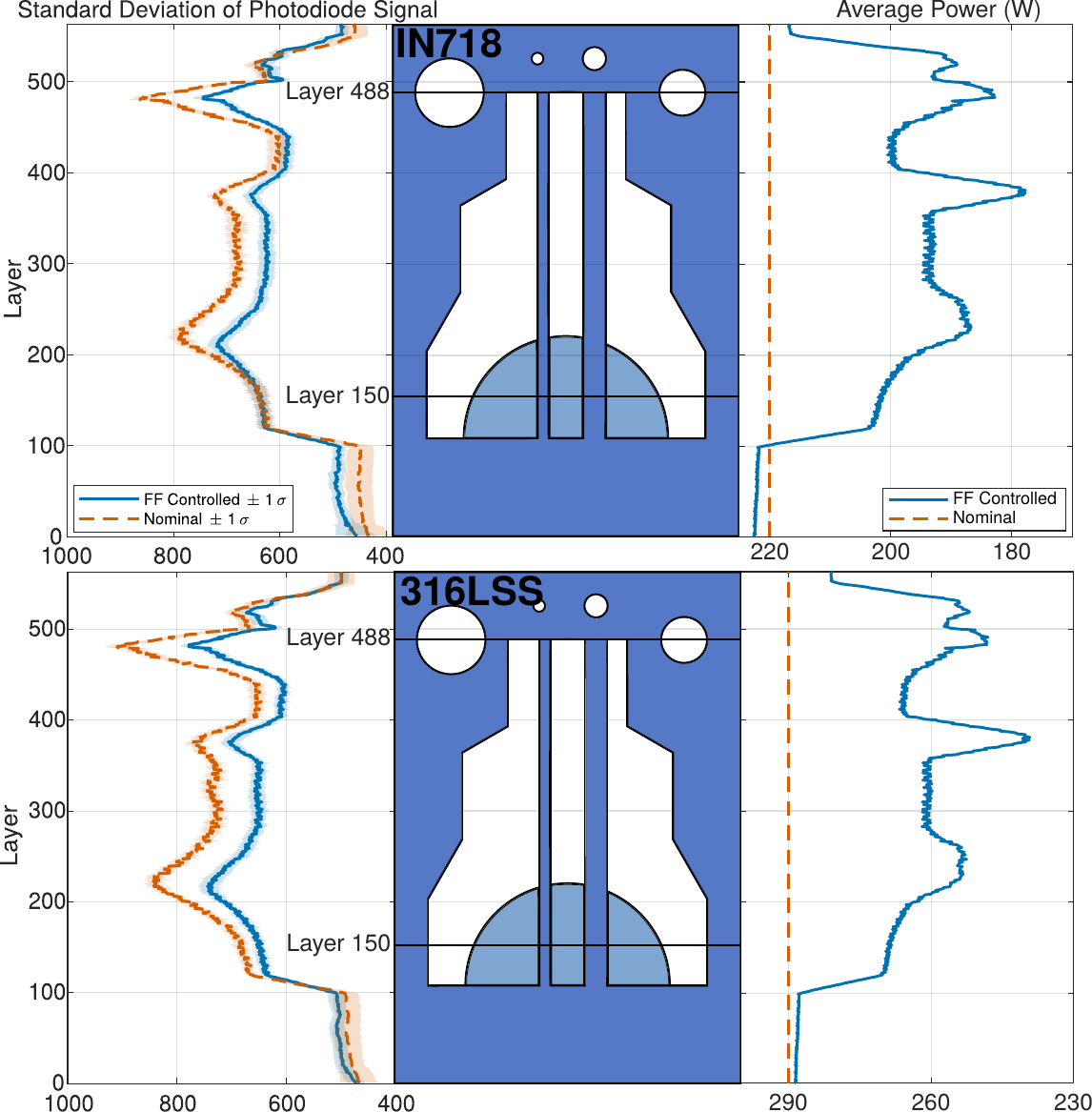}
    \caption{The standard deviation of the photodiode signal in each layer (unitless), averaged over three samples, and the average power in each layer for both IN718 (top) and 316LSS (bottom). The colored bar represents \(\pm 1\) standard deviation of the three printed samples. A 20-layer moving average has been applied to improve readability and reduce noise.}
    \label{fig:photodiodeVariation_and_power_layerwise}
\end{figure*}

In the following subsection, a closer analysis is provided for layers 150 and 488, labeled on Figure~\ref{fig:photodiodeVariation_and_power_layerwise}, which highlights behaviors in key features of the part. Additionally, an analysis of the geometric accuracy and porosity will be presented.

\subsection{Local Analysis of Key Layers}\label{sec:key_layer_results}

\subsubsection{Intralayer Scheduled Power}

\begin{figure*}
    \centering
    \includegraphics[width=\linewidth]{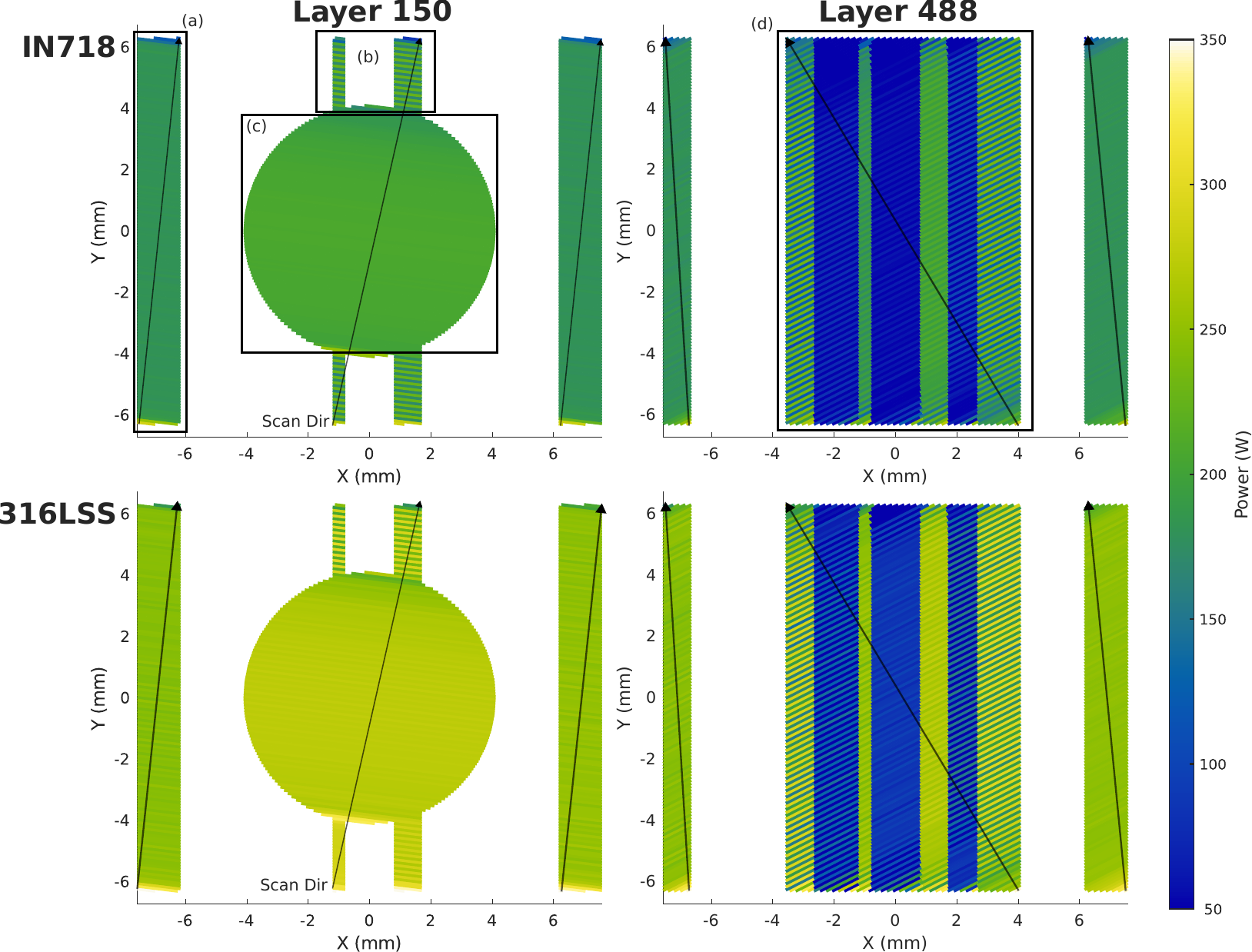}
    \caption{The scheduled laser power by the controller for two layers of the test geometry in IN718 (top) and 316LSS (bottom). Layer locations are indicated on Figure~\ref{fig:cologne_bottle_dims}. Key regions are labeled on IN718 as follows: (a) pillar, (b) thin wall structures, (c) solid part, (d) the overhang region.}
    \label{fig:Laser_power}
\end{figure*}

Inspecting the laser power within a layer, rather than simply the average power, showcases how the model compensates for varying vector lengths and temporal effects. Both materials behave similarly, and four key regions are examined as labeled in Figure~\ref{fig:Laser_power}.

Examining region (a) in layer 150, the pillar, the scan starts with a higher power at the beginning of the scan before dropping slightly below nominal power, then dropping sharply at the end of the region. (Note that the direction of travel from bottom left to top right is shown by the arrow.) This makes sense, as at the beginning, the part has not heated up; however, within a few vectors, it converges to a steady-state subsurface temperature. The shorter vectors at the end, combined with the region having nowhere for heat to dissipate, cause the subsurface temperature to rise. This results in \(P^*(l_v)\) being high initially, converging to a steady state, and then dropping at the end of the pillar. 

Examining region (b), thin wall structures, we see similar behavior to the pillar region but with some oscillations. This is due to the scanpath going across both thin walls in one pass. The laser makes one full pass across both thin walls. Upon turning around, it immediately reheats the same wall and so drops the power; however, by the time it reaches the other wall, the wall has cooled, and so it increases the power. This repeats, creating the observed oscillations.

In the solid hemisphere in the center of the part, region (c), the controller has a similar power to the nominal case, as heat dissipates easily in this region. However, the beginning and end of this region behave similarly to the beginning and end of the pillars.

In layer 488, examining region (d), the overhang region, two interesting behaviors are observed. First, the controller chooses a significantly lower power in the actual overhangs. The FDM thermal model captures the severe overheating that can be seen in both the photodiode signal in Figure~\ref{fig:layer_488_photodidoe_signal} and in "IN718\_Thermal\_Camera.mp4" and "316LSS\_Thermal\_Camera.mp4" in the supplemental material. Second, there are more exaggerated oscillations at the turnaround locations. Similarly to region (b), the controller compensates for increased heating. The algorithm divides the longer vectors into multiple short ones at the overhangs, requiring skywriting to be turned off in this region. Hence, the laser turns around immediately instead of skywriting, and the controller partially compensates for this by dropping the power.

These power profiles illustrate how the controller tailors input to local geometry, particularly near overhangs, aligning with the hypothesis that feedforward methods can proactively mitigate thermal accumulation and regulate melt pool dimensions. Overall, both materials respond similarly, with the main difference between the two being the scaling of the power input.

\subsubsection{Intralayer Photodiode Signal}

\begin{figure*}
    \centering
    \includegraphics[width=\linewidth]{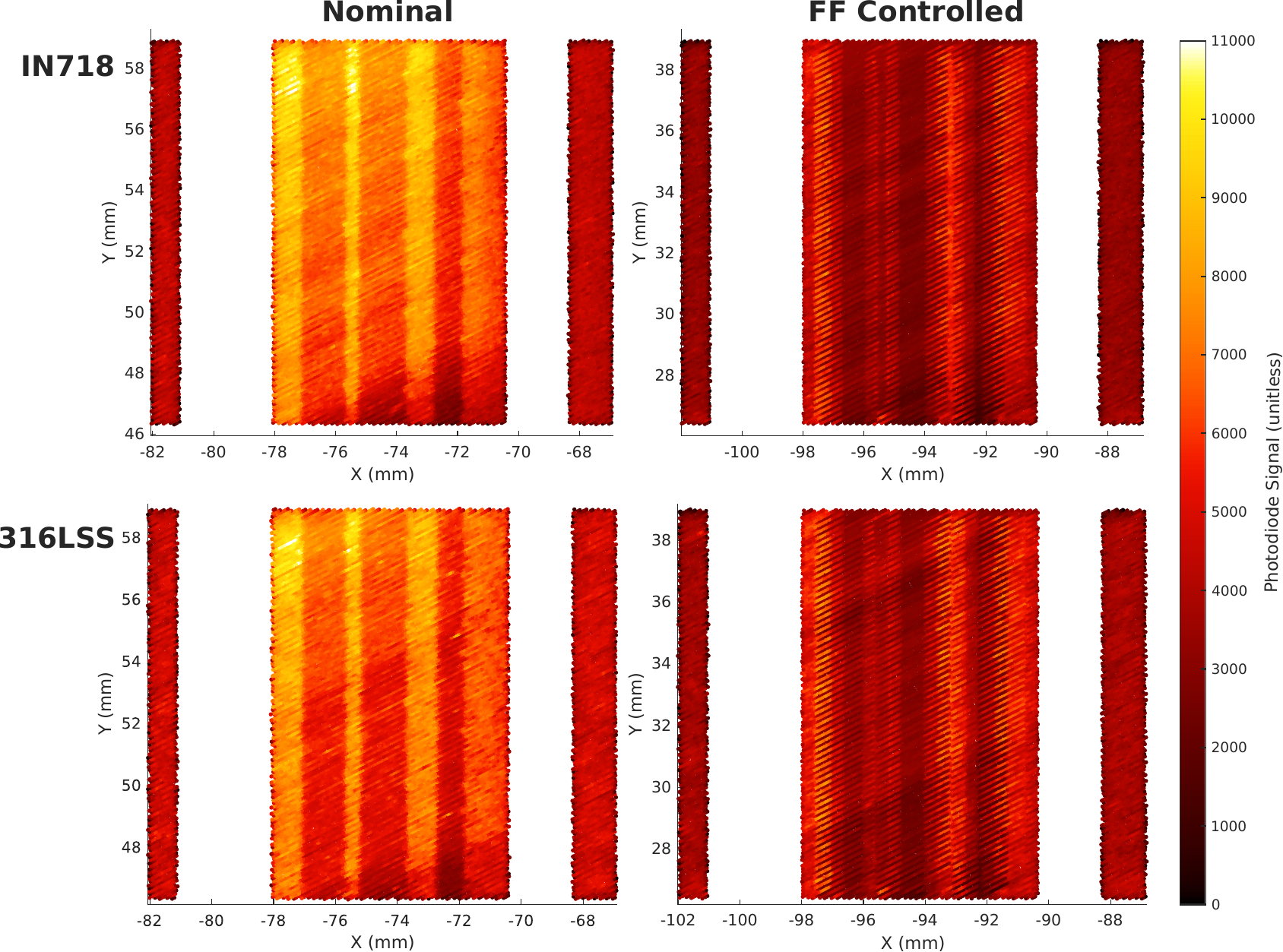}
    \caption{The resulting photodiode signal for both the Nominal (left) and FF Controlled (right) in IN718 (top) and 316LSS (bottom) in layer 488. The location of layer 488 on the part can be seen in Figure~\ref{fig:photodiodeVariation_and_power_layerwise}. Positions are given relative to the baseplate.}
    \label{fig:layer_488_photodidoe_signal}
\end{figure*}

Given the significant change in commanded power in region (d), layer 488 was chosen for further investigation into the photodiode signal. Figure~\ref{fig:layer_488_photodidoe_signal} shows the photodiode signal for both the Nominal and FF Controlled prints in both materials. The nominal sample had high variation (\(\sigma=1852\)) and an overall high signal intensity. This implies an unstable melt pool and likely overheating. In contrast, the controlled signal showed a \(38\%\) lower variation (\(\sigma=1157\)) and an overall lower intensity. The lower variation indicates a more consistent melt pool across the part, and the lower intensity implies less overall heating. "IN718\_Thermal\_Camera.mp4" and "316LSS\_Thermal\_Camera.mp4" in the supplemental material supported this observation.

The improvement in signal variation and intensity in layer 488 suggests that the controller, by dropping the power, successfully decreased in-layer melt-pool variation, increased melt-pool stability, and provided a more uniform melt-pool size. This aligns with the layer-wise observations that power adjustment decreased the photodiode signal variance in geometrically complex regions.

\subsection{Geometric Results}

Critical dimensions, given in Figure~\ref{fig:cologne_bottle_dims} for both the Nominal and FF Controlled samples in both materials, were recorded in Table \ref{tbl:critical_dims_geometry}. Visuals of these regions for IN718 can be seen in the optical microscope data in Figure~\ref{fig:inconel_opt_microscope}.

Significant inaccuracies were observed in the Nominal samples for both materials, where overheating in the overhangs suggested that the melt pool enlarged or even transitioned to keyhole mode, which caused significant dross formation. This overheating also reduced the hole dimensions, decreasing the vertical dimensions from the as-designed value by an average of \(17\%\). Comparatively, the FF Controlled samples exhibited significantly less geometric error, with reduced dross in overhang regions and improved dimensional accuracy in the holes, which resulted in an average error of \(11\%\). The drop in power in the overhang region, as seen in Figure~\ref{fig:Laser_power}, region (d), and the lower power in other overhang regions, significantly decreased the overheating and thus indicates that the melt pool is a more consistent size. These more consistent melt pool dimensions resulted in improved geometric accuracy.

\begin{table*}[h!]
\centering
\caption{Comparison of the key dimensions of the cologne bottle for Nominal and FF Controlled samples in \textit{IN718} and \textit{316LSS} as measured by X-ray CT scan.}
\begin{tabular}{m{1cm} m{1.5cm} m{2.25cm} m{2.5cm} m{2.25cm} m{2.5cm}}
    \toprule
    &  & \multicolumn{2}{c}{\textbf{IN718}} & \multicolumn{2}{c}{\textbf{316LSS}}\\
    \cmidrule(lr){3-4}\cmidrule(lr){5-6}
    & Design (\(mm\)) & Nominal (\%~error) & FF \linebreak Controlled (\%~error) & Nominal (\%~error) & FF \linebreak Controlled (\%~error)\\
    \midrule
    D1 & 3    & 2.75 (8.3\%)  & 2.89 (3.5\%)  & 2.72 (9.5\%)  & 2.85 (5.0\%)\\
    D2 & 2    & 1.73 (14\%)   & 1.82 (9.0\%)  & 1.64 (18\%)   & 1.79 (11\%)\\
    D3 & 1    & 0.73 (27\%)   & 0.83 (17\%)   & 0.64 (36\%)   & 0.81 (19\%)\\
    H1 & 15.2 & 14.74 (3\%)   & 15.23 (0.2\%)& 14.88 (2\%)   & 15.16 (0.3\%)\\
    H2 & 15.2 & 14.96 (1.5\%) & 15.25 (0.3\%)& 14.85 (2\%)   & 15.12 (0.5\%)\\
    \bottomrule
\end{tabular}
\label{tbl:critical_dims_geometry}
\end{table*}

\subsubsection{Porosity}

\begin{table}
\caption{Porosity metrics obtained from optical microscopy for \textit{IN718} and \textit{316LSS}. Pores are computed by thresholding of the images as described in Section~\ref{sec:validation_test_setup}.}
\label{tbl:porosity_metrics}
\small

\begin{subtable}{\linewidth}
\subcaption{Overhang Region}
\begin{tabular*}{\tblwidth}{@{}lccc@{}}
\toprule
\multirow{2}{*}{Material} & \multicolumn{2}{c}{Density (\%)} &
  \multirow{2}{*}{Reduction in Porosity (\%)}\\
\cmidrule(lr){2-3}
  & Nom. & FF & \\  %
\midrule
IN718  & 99.56 & 99.83 & 61\\
316LSS & 99.71 & 99.77 & 24\\
\bottomrule
\end{tabular*}
\end{subtable}

\vspace{8pt}

\begin{subtable}{\linewidth}
\subcaption{Whole-part}
\begin{tabular*}{\tblwidth}{@{}lccc@{}}
\toprule
\multirow{2}{*}{Material} & \multicolumn{2}{c}{Density (\%)} &
  \multirow{2}{*}{Reduction in Porosity (\%)}\\
\cmidrule(lr){2-3}
  & Nom. & FF & \\  %
\midrule
IN718  & 99.71 & 99.76 & 17\\
316LSS & 99.69 & 99.74 & 16\\
\bottomrule
\end{tabular*}
\end{subtable}
\vspace{8pt}

\begin{subtable}{\linewidth}
\subcaption{Pores \(<\!6\,\mu m\) in equivalent diameter (\% of total)}
\begin{tabular*}{\tblwidth}{@{}lcc@{}}
\toprule
Material & Nom. & FF\\
\midrule
IN718  & 82 & 66\\
316LSS & 62 & 61 \\
\bottomrule
\end{tabular*}
\end{subtable}
\end{table}

Table~\ref{tbl:porosity_metrics} summarizes the porosity metrics extracted from the optical micrographs. For IN718, the FF controller raised the overhang-region density from 99.56\% to 99.83\%, which eliminated roughly 61\% of the local pore cross-sectional area. A less significant 17\% reduction was achieved for the whole part, driven by a decrease in sub-\(6\,\mu m\) pores (82\% down to 66\% of the total).

The same pattern appears in 316LSS, albeit with smaller absolute changes: overhang density improved from 99.71\% to 99.77\%, a 24\% reduction in pore cross-sectional area, and whole-part density increased from 99.69\% to 99.74\%, a 16\% reduction in porosity. In this material, the controller mainly suppressed larger pores, as the share of sub-\(6\,\mu m\) voids remained essentially unchanged (62–61\%).

\begin{figure}
    \centering
    \includegraphics[width=0.5\linewidth]{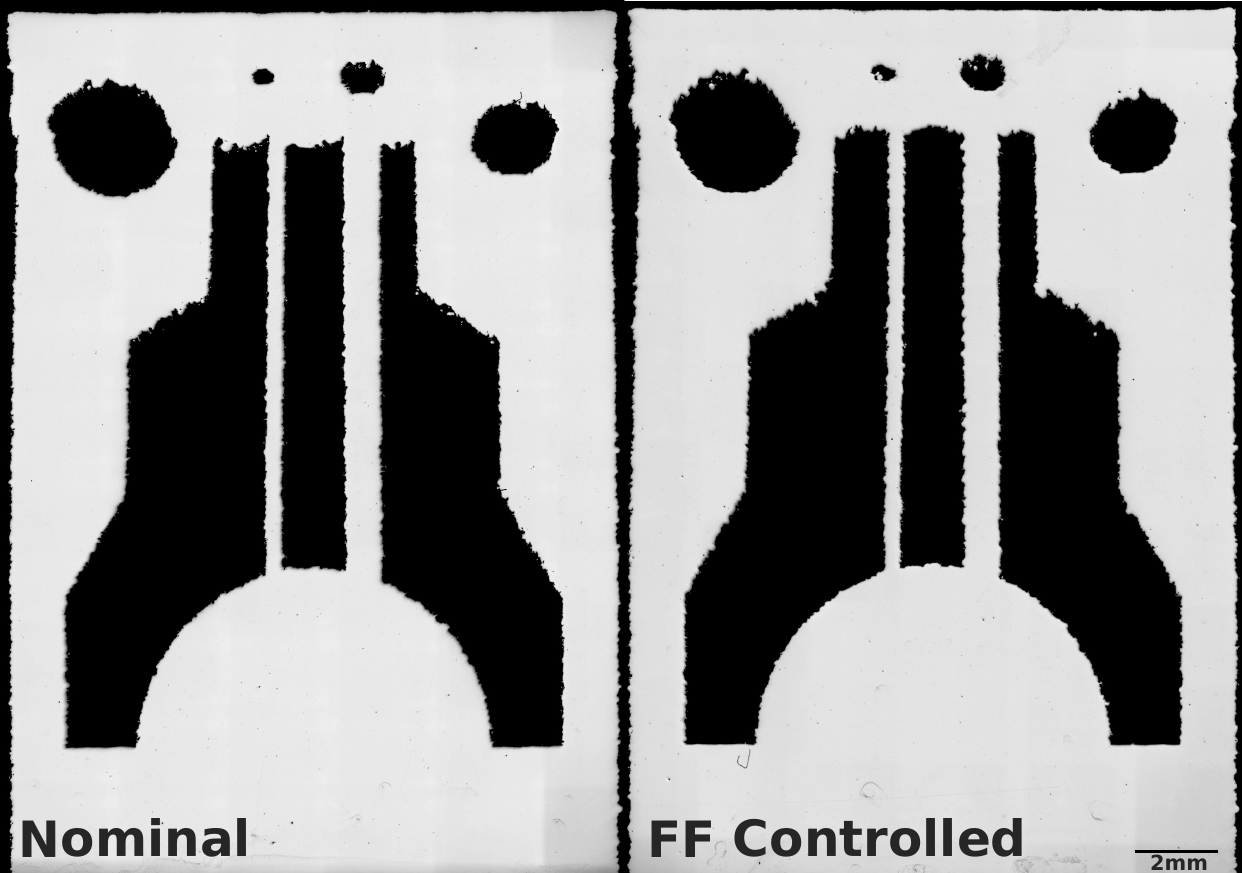}
    \caption{Optical images of the polished IN718 samples. Nominal (left) and FF Controlled (right). The dross in key overhang regions is visibly reduced in the FF Controlled case.}
    \label{fig:inconel_opt_microscope}
\end{figure}

Together with the visibly cleaner overhangs in the optical microscope Figure~\ref{fig:inconel_opt_microscope}, this reduction in porosity across the part indicates that the controller is mitigating overheating and occurrences where the melt pool enters the keyholing regime. However, the significant drop in porosity \(< 6\mu m\) in equivalent diameter for IN718 is unlikely to be due to a reduction in keyholing, as the pores are smaller than previously observed for keyholing \cite{huangKeyholeFluctuationPore2022}. We do not currently understand the reason behind the reduction of these smaller pores by the FF controller.

\section{Conclusion}\label{text_conclusion}

This paper has introduced a novel feedforward control framework that decouples part-scale thermal effects from small-scale melt pool dynamics. This decoupling enabled scale-agnostic melt pool prediction and efficient optimization over the melt pool dimensions. To implement this framework, an analytical melt pool model, dependent solely on subsurface temperature, speed, and power, and a lightweight, coarse FDM thermal model capable of simulating the vector-level thermal behavior for part-scale objects were derived. Each was calibrated separately using single-track and 2D scans, respectively, and then combined to generate feedforward laser power schedules at the part scale across two separate materials, Inconel 718 and 316L stainless steel, demonstrating that a fast, part-scale thermal and melt pool model, coupled with straightforward calibration on bare-metal plates, can significantly enhance the quality of LPBF parts. Although validated only on a single machine setup and two materials, the framework is readily extensible to other platforms and alloys, as long as a simple process map has been developed.

We have shown the improvements resulting from the vector-level feedforward controller in experimental studies on two materials, 316LSS and IN718, leading to the following conclusions:

\begin{enumerate}
    \item[(1)] Decoupling melt pool dynamics from part-scale thermal fields through the physics-based coupled analytical-numerical model enabled a prediction of the melt pool area at part-scale while requiring limited calibration on 2D plates only.
    \item[(2)] Using this method to preemptively modulate the power on a vector-by-vector level, compensated for local thermal buildup, and maintained a more consistent melt pool. This was demonstrated by an average decrease in photodiode variation of \(5.3\%\) for IN718 and \(8.3\%\) for 316LSS in layers 100-550 of the evaluated part, which are prone to thermal buildup.
    \item[(3)] Reduced variations in melt pool area, especially in overhang layers, resulted in up to a \(62\%\) improvement on average in geometric accuracy of key dimensions.
    \item[(4)] In conjunction, up to \(42.5\%\) reduction in porosity was observed in key overhang regions with up to overall \(16.5\%\) decrease in porosity across the part.
\end{enumerate}

The first aspect of future work we identify is the need for fast, reduced-order models between process signatures, such as the melt pool area, and structural or engineering features of interest. The constant area target used in this work may not correlate with achieving the desired engineering performance metrics, such as geometric accuracy or minimal porosity. While the controller can track a spatiotemporally varying \(A_c(x,y,z,t)\) to achieve the desired properties, to our knowledge, there is a lack of models for generating such distributions. As such, a mapping between these complex process outcomes and the melt pool area, or other melt pool features, needs to be developed. This would enable the controller to target these engineering performance metrics better.

Another aspect of future work is addressing existing limitations with the part-scale thermal model: incorporating numerical methods that make accuracy less dependent on an everywhere-dense grid, developing more sophisticated ways of modeling conduction into the powder, developing better models for the interlayer dwell, and incorporating data-driven elements into the model to capture nonlinear modes of heat transfer without undue computational burden. Further experiments can be conducted to quantify the controller's effect on material properties and microstructure. 

\section*{CRediT authorship contribution statement}\label{sec: CRediT}
\textbf{Nicholas Kirschbaum:} Conceptualization, Methodology, Software, Formal analysis, Validation, Data curation,
Writing - original draft, Writing - review \& editing. \textbf{Nathaniel
Wood:} Conceptualization, Methodology, Software, Writing
- review \& editing. \textbf{Chang-Eun Kim:} Methodology, Resources, Writing - review \& editing. \textbf{Thejaswi Tumkur:} Resources, Writing - review \& editing, Project administration, Funding Acquisition. \textbf{Chinedum Okwudire:}  Conceptualization, Methodology, Validation, Writing - review \& editing, Supervision, Funding acquisition.

\section*{Declaration of Generative AI and AI-assisted technologies in the writing process.}
During the preparation of this work, the author(s) used ChatGPT (OpenAI) to check for grammar errors and improve the language and readability of the manuscript. After using this tool, the author(s) reviewed and edited the content as needed and take(s) full responsibility for the content of the published article.

\section*{Acknowledgment}
This work was performed under the auspices of the U.S. Department of Energy by Lawrence Livermore National Laboratory (LLNL) under Contract DE-AC52-07NA27344. We thank Mr. Gabe Guss and other members affiliated with the Materials Science Division at LLNL for their valuable feedback on this work. LLNL-JRNL-2007913

\section*{Declaration of Competing Interest}
The authors declare that they have no known competing financial interests or personal relationships that could have appeared to influence the work reported in this paper.

\section*{Data Availability}
The authors will make any data and computer code from this work available upon reasonable request.

\appendix

\section*{Appendices}

\section{Interlayer Dwell Computation}\label{sec:interlayer_dwell}

Accounting for the thermal dissipation during the interlayer dwell is crucial to maintaining an approximately correct thermal field. However, the element density is such that explicitly simulating it is computationally prohibitive. As such, a combination of 1D analytical solutions and 2D Gaussian blurs is used.

\begin{assumption}
    Heat transfer is predominantly in the \(Z\) direction. Thus, over long periods of time, parallel 1D heat transfer problems approximate the real temperature field.
\end{assumption}

\begin{assumption}
    Due to the shape of the Green's function for the heat conduction equation, intra-layer heat transfer can be approximated via a Gaussian blur as done elsewhere in the literature \cite{wolferFastSolutionStrategy2019}.
\end{assumption}

\begin{assumption}
    Over the comparatively extreme timescales of the interlayer dwell, heat spreads out sufficiently that artifacts introduced by the two-stage \(Z\)-then-\(XY\) approximation of the thermal dissipation are minimal.
\end{assumption}

\begin{figure}
    \centering
    \includegraphics[width=0.6\linewidth]{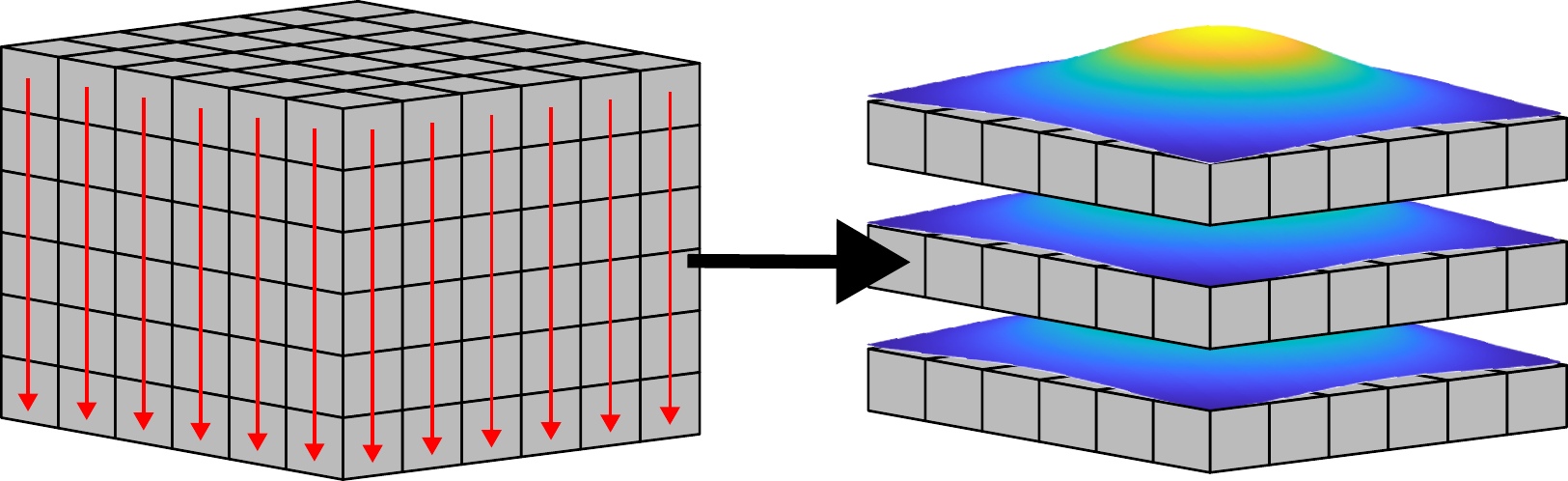}
    \caption{To perform the interlayer dwell, a series of 1D analytical heat transfer problems are first solved to propagate heat in \(Z\). Then, a 2D Gaussian blur is applied to propagate remaining heat in \(X\) and \(Y\).}
    \label{fig:interlayer_dwell_steps}
\end{figure}

First, the explicitly simulated layers, used to determine \(P^*(l_v)\), are mapped onto a discretization of the full part. Because there is uniform grid spacing in the XY-plane, elements can be grouped in the Z-direction along ``lines'' of constant $(i,j)$ index. This enables each line of Z elements to be mapped to a 1D analytical heat transfer problem, the four cases of which can be seen in Figure~\ref{fig:four_cases}. This is used to perform the first step of the dwell shown in Figure~\ref{fig:interlayer_dwell_steps}. This enables jumping to the end of the dwell in one step. 

Once the 1D analytical heat transfer is complete for all lines of Z, a Gaussian blur is applied to capture the lateral heat diffusion. To accomplish this, the temperature field for each layer is mapped to an image where the powder is represented by a pixel at half the average temperature of the layer post 1D heat transfer. This enables heat leakage into the powder without overly dropping the temperature of the part during the interlayer dwell. Then, a Gaussian blur with a standard deviation of \(\sigma=\frac{\sqrt{2\alpha \Delta t}}{h_s}\) \cite{wolferFastSolutionStrategy2019} is applied to each layer to simulate a 2D heat transfer of time \(\Delta t\). This approximates 2D heat transfer by setting the standard deviation equal to one characteristic thermal diffusion length, thereby preventing heat entrapment in the 1D insulation-to-insulation case.

Once both steps are completed, the temperature field is mapped back onto the explicitly simulated layers. The new layer (representing powder being spread) is assumed to be at half the temperature of the underlying material.

The derivation of the 1D analytical solutions for the four cases as applied to the discretized space is given in the following sections.

\subsection{Initial Condition Formulation}
Given the nodal values \(T_1, T_2, \ldots, T_N\) of the \(N\) elements in a line in the z direction (see Figure~\ref{fig:four_cases}), with coordinates \(z_1, z_2, \ldots, z_N\) and total length \(L\), we approximate the continuous temperature field spanning them as a piecewise linear function.

We divide the span of \(N\) elements into \((N-1)\) regions. The \(e\)th region is bounded by the values \((z_1^e, z_2^e)\), and has the temperatures \((T_1^e, T_2^e)\) at these points. Define
\begin{equation}\label{eq:1D_init_cond}
T_{ic}(z) \;=\; \sum_{e=1}^{N-1} f_e (z),
\end{equation}
where
\[
f_e(z) \,=\,
\begin{cases}
\displaystyle
\frac{T_2^e - T_1^e}{z_2^e - z_1^e} \,\bigl(z - z_1^e\bigr) \;+\; T_1^e, 
& z \in (z_1^e,\, z_2^e), \\[6pt]
0, & \text{otherwise}.
\end{cases}
\]
This has now derived the initial temperature field that will be used in the 1D heat transfer.
\subsection{1D Heat Transfer Problem}
\begin{figure}
    \centering
    \includegraphics[width=0.4\linewidth]{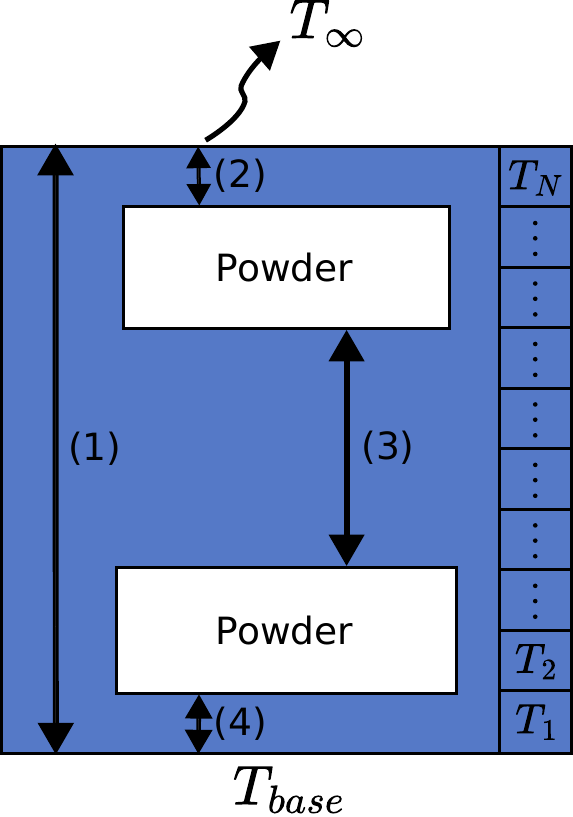}
    \caption{The four cases for the 1D analytical heat transfer. (1) Convection to Constant Temperature. (2) Convection to Insulation. (3) Insulation to Insulation. (4) Insulation to Constant Temperature.}
    \label{fig:four_cases}
\end{figure}

We formulate the 1D heat transfer as four possible cases, as shown in Figure~\ref{fig:four_cases}.

\begin{enumerate}
    \item[(1)] Convection to Constant Temperature
    \item[(2)] Convection to Insulation
    \item[(3)] Insulation to Insulation
    \item[(4)] Insulation to Constant Temperature
\end{enumerate}
These track the four distinct lines of z elements found in the part: the top of the part to the bottom, the top to powder, powder to powder, and finally, powder to bottom. Each solution solves the 1D heat transfer problem with varying boundary conditions given the initial condition in Eq.~\eqref{eq:1D_init_cond}. A summary of these problems and their general solutions can be found in Table \ref{table:summary_1D}. The following sections compute the specific coefficients for each case.

\begin{sidewaystable}[h!]
\centering
\renewcommand{\arraystretch}{1.25}
\begin{tabular*}{1.07\tblwidth}{@{}clll@{}}
\toprule
Case & Boundary conditions & Eigenvalue rule & Closed–form solution (with $C_n$ from projection)\\ \midrule
(1) &
\makecell{$T(0,t)=T_0$\\$-k\,T_z(L,t)=h\,[T(L,t)-T_\infty]$} &
$\displaystyle\lambda_n\cot(\lambda_n L)+\frac{h}{k}=0$ &
\makecell[l]{$T(z,t)=Kz+T_0+\sum_{n}C_n e^{-\lambda_n^2\alpha t}\sin(\lambda_n z)$\\[2pt]
$K=\dfrac{h(T_\infty-T_0)}{hL+k}$\\
$\displaystyle C_n=\frac{2\int_0^{L}[T_{ic}-T_0-Kz]\,\sin(\lambda_n z)\,dz}{\,L-\frac{1}{2\lambda_n}\sin(2\lambda_n L)}$} \\ \midrule
(2) &
\makecell{$T_z(0,t)=0$\\$-k\,T_z(L,t)=h\,[T(L,t)-T_\infty]$} &
$\displaystyle\lambda_n\tan(\lambda_n L)-\frac{h}{k}=0$ &
\makecell[l]{$T(z,t)=T_\infty+\sum_{n}C_n e^{-\lambda_n^2\alpha t}\cos(\lambda_n z)$\\[2pt]
$\displaystyle C_n=\frac{2\int_0^{L}[T_{ic}-T_\infty]\cos(\lambda_n z)\,dz}{\,L+\frac{1}{2\lambda_n}\sin(2\lambda_n L)}$} \\ \midrule
(3) &
\makecell{$T_z(0,t)=0$\\$T_z(L,t)=0$} &
$\displaystyle\lambda_n=\frac{n\pi}{L}$ &
\makecell[l]{$T(z,t)=C_0+\sum_{n\ge1}C_n e^{-\lambda_n^2\alpha t}\cos(\lambda_n z)$\\[2pt]
$C_0=\dfrac{1}{L}\int_0^{L}T_{ic}\,dz$,\quad
$\displaystyle C_n=\frac{2}{L}\int_0^{L}[T_{ic}-C_0]\cos(\lambda_n z)\,dz$} \\ \midrule
(4)&
\makecell{$T(0,t)=T_0$\\$T_z(L,t)=0$} &
$\displaystyle\lambda_{n}=\frac{(2n+1)\pi}{2L}$ &
\makecell[l]{$T(z,t)=T_0+\sum_{n}C_n e^{-\lambda_n^2\alpha t}\sin(\lambda_n z)$\\[2pt]
$\displaystyle C_n=\frac{2}{L}\int_0^{L}[T_{ic}-T_0]\sin(\lambda_n z)\,dz$} \\ \bottomrule
\end{tabular*}
\caption{Summary of 1-D transient-conduction solutions for the four boundary-condition combinations.  Each coefficient is found from a single projection integral of the (piecewise-linear) initial temperature profile.}
\label{table:summary_1D}
\end{sidewaystable}

\subsubsection{Case 1: Convection to Constant Temperature}

As \(T_{ic}(z)\) is piecewise linear \(C_n\) reduces to:
\begin{equation}
C_{n} =
  \frac{2%
        \displaystyle\sum_{e=1}^{N-1}\!
        \int_{z_{1}^{e}}^{z_{2}^{e}}
          \Bigl[
            \bigl(\tfrac{T_{2}^{e}-T_{1}^{e}}{z_{2}^{e}-z_{1}^{e}}\bigr)z
            +T_{1}^{e}-T_{0}-Kz_{1}^{e}
          \bigr]\sin\!\bigl(\lambda_{n}z\bigr)\,dz}
       {L-\tfrac{1}{2\lambda_{n}}\sin(2\lambda_{n}L)}
\end{equation}

This can then be solved to determine \(C_n\), which, combined with the general solution in Table \ref{table:summary_1D}, gives the specific solution.
\begin{equation}
\begin{aligned}
C_{n} &= \left(\frac{2}{ L - \frac{1}{2 \lambda_{n}}\sin\bigl(2 \lambda_{n} L\bigr)}\right)
\sum_{e=1}^{N-1}
\Biggl[
  C_{0}^{e}\Bigl(
    \frac{z_{1}^{e}\cos(\lambda_{n} z_{1}^{e}) \;-\; z_{2}^{e}\cos(\lambda_{n} z_{2}^{e})}{\lambda_{n}}
    \\&+
    \frac{\sin(\lambda_{n} z_{2}^{e}) \;-\; \sin(\lambda_{n} z_{1}^{e})}{\lambda_{n}^{2}}
  \Bigr)
  +\;
  C_{1}^{e}\Bigl(
    \frac{\cos(\lambda_{n} z_{1}^{e}) \;-\; \cos(\lambda_{n} z_{2}^{e})}{\lambda_{n}}
  \Bigr)
\Biggr],\\[6pt]
C_{0}^{e} &= \frac{T_{2}^{e} - T_{1}^{e}}{z_{2}^{e} - z_{1}^{e}} \;-\; K,\\[6pt]
C_{1}^{e} &= T_{1}^{e}
            \;-\;
            \Bigl(\frac{T_{2}^{e} - T_{1}^{e}}{z_{2}^{e} - z_{1}^{e}}\Bigr) z_{1}^{e}
            \;-\;
            T_{0}
\end{aligned}
\end{equation}

\subsubsection{Case 2: Convection to Insulation}

As \(T_{ic}(z)\) is piecewise linear \(C_n\) reduces to:
\begin{equation}
C_{n} =
  \frac{2
        \displaystyle\sum_{e=1}^{N-1}\!
        \int_{z_{1}^{e}}^{z_{2}^{e}}
          \Bigl[
            \bigl(\tfrac{T_{2}^{e}-T_{1}^{e}}{z_{2}^{e}-z_{1}^{e}}\bigr)
            (z-z_{1}^{e})+T_{1}^{e}-T_{\infty}
          \Bigr]\cos\!\bigl(\lambda_{n}z\bigr)\,dz}
       {L+\tfrac{1}{2\lambda_{n}}\sin(2\lambda_{n}L)}
\end{equation}

This can then be solved to determine \(C_n\), which, combined with the general solution in Table \ref{table:summary_1D}, gives the specific solution.
\begin{equation}
\begin{aligned}
C_{n}&=\left(
    \frac{2}{ 
       L-\frac{1}{2\lambda_{n}}\sin\bigl(2\lambda_{n}L\bigr)}
    \right)
\sum_{e=1}^{N-1}
\Biggl[
  C_{0}^{e}\Bigl(
      \frac{z_{2}^{e}\sin(\lambda_{n}z_{2}^{e})
            -z_{1}^{e}\sin(\lambda_{n}z_{1}^{e})}{\lambda_{n}}\\
      &+\frac{\cos(\lambda_{n}z_{2}^{e})
             -\cos(\lambda_{n}z_{1}^{e})}{\lambda_{n}^{2}}  \Bigr)
  + C_{1}^{e}\Bigl(
      \frac{\sin(\lambda_{n}z_{2}^{e})
            -\sin(\lambda_{n}z_{1}^{e})}{\lambda_{n}^{2}}
  \Bigr)
\Biggr],\\[8pt]
C_{0}^{e}&=\frac{T_{2}^{e}-T_{1}^{e}}{z_{2}^{e}-z_{1}^{e}},\\[6pt]
C_{1}^{e}&=T_{1}^{e}
           -\Bigl(\frac{T_{2}^{e}-T_{1}^{e}}{z_{2}^{e}-z_{1}^{e}}\Bigr)z_{1}^{e}
           -T_{\infty}.
\end{aligned}
\end{equation}

\subsubsection{Case 3: Insulation to Insulation}

As \(T_{ic}(z)\) is piecewise linear \(C_0\) reduces to:
\begin{equation}\label{eq:C0_int}
C_{0}=\frac{1}{L}
        \sum_{e=1}^{N-1}\Bigl[
          \tfrac{(T_{2}^{e}-T_{1}^{e})(z_{2}^{e}+z_{1}^{e})}{2}
          +T_{1}^{e}(z_{2}^{e}-z_{1}^{e})
          -(T_{2}^{e}-T_{1}^{e})z_{1}^{e}
        \Bigr]
\end{equation}

As \(T_{ic}(z)\) is piecewise linear \(C_n\) reduces to:
\begin{equation}\label{eq:piecewise_int}
C_{n}=
  \frac{2}{L}%
  \sum_{e=1}^{N-1}\!
  \int_{z_{1}^{e}}^{z_{2}^{e}}
    \Bigl[
      \tfrac{T_{2}^{e}-T_{1}^{e}}{z_{2}^{e}-z_{1}^{e}}(z-z_{1}^{e})
      +T_{1}^{e}
    \Bigr]\cos\!\bigl(\lambda_{n}z\bigr)dz
\end{equation}

This can then be solved to determine \(C_n\), which, combined with the general solution in Table \ref{table:summary_1D}, gives the specific solution.
\begin{equation}\label{eq:Ck_final}
\begin{aligned}
C_{n}&= \frac{2}{L}
       \sum_{e=1}^{N-1}
       \Biggl[
         C_{0}^{e}\Bigl(
           \frac{z_{2}^{e}\sin(\lambda_{n}z_{2}^{e})
                 -z_{1}^{e}\sin(\lambda_{n}z_{1}^{e})}{\lambda_{n}}\\
           &+\frac{\cos(\lambda_{n}z_{2}^{e})
                  -\cos(\lambda_{n}z_{1}^{e})}{\lambda_{n}^{2}}
         \Bigr)
         +C_{1}^{e}
          \Bigl(
            \frac{\sin(\lambda_{n}z_{2}^{e})
                  -\sin(\lambda_{n}z_{1}^{e})}{\lambda_{n}}
          \Bigr)
       \Biggr]\\[4pt]
       C_{0}^{e}&= \frac{T_{2}^{e}-T_{1}^{e}}{z_{2}^{e}-z_{1}^{e}}\\[4pt]
C_{1}^{e}&= T_1^e-\frac{{T_2^e-T_1^e}}{z_2^e-z_1^e}z_1^e
\end{aligned}
\end{equation}

\subsubsection{Case 4: Insulation to Constant Temperature}

As \(T_{ic}(z)\) is piecewise linear \(C_n\) reduces to:
\begin{equation}
C_{n}=
  \frac{2}{L}%
  \sum_{e=1}^{N-1}\!
    \int_{z_{1}^{e}}^{z_{2}^{e}}
      \Bigl[
        \tfrac{T_{2}^{e}-T_{1}^{e}}{z_{2}^{e}-z_{1}^{e}}(z-z_{1}^{e})
        +T_{1}^{e}-T_{0}
      \Bigr]\sin\!\bigl(\lambda_{n}z\bigr)dz
\end{equation}

This can then be solved to determine \(C_n\), which, combined with the general solution in Table \ref{table:summary_1D}, gives the specific solution.
\begin{equation}
\begin{split}
C_n &= \frac{2}{L} \sum_{e=1}^{N-1}
\Biggl[
  C_0^e \left(
    \frac{z_1^e \cos(\lambda_{n} z_1^e) - z_2^e \cos(\lambda_{n} z_2^e)}{\lambda_{n}}\right.\\
    & \left.+ \frac{\sin(\lambda_{n} z_2^e) - \sin(\lambda_{n} z_1^e)}{\lambda_{n}^2}
  \right) 
  + C_1^e \left(
    \frac{\cos(\lambda_{n} z_1^e) - \cos(\lambda_{n} z_2^e)}{\lambda_{n}}
  \right)
\Biggr], \\[6pt]
C_0^e &= \frac{T_2^e - T_1^e}{z_2^e - z_1^e}, \\[6pt]
C_1^e &= T_1^e - \left(\frac{T_2^e - T_1^e}{z_2^e - z_1^e}\right)z_1^e - T_0
\end{split}
\end{equation}

\bibliographystyle{elsarticle-num.bst}

\bibliography{Vector_Level_Control_Paper.bib}

\end{document}